\documentclass[pre,showpacs,twocolumn,amsmath,amssymb,showpacs]{revtex4}
\usepackage{graphicx}% Include figure files
\usepackage{dcolumn}% Align table columns on decimal point
\usepackage{bm}% bold math
\newcommand{\bfa}{\mathbf{a}}
\newcommand{\bfx}{\mathbf{x}}
\newcommand{\bfeta}{\mbox{\boldmath $\bf\eta$}}
\newcommand{\nus}[1]{\nu_{\!{}_{#1}}}
\newcommand{\spaEq}{\hspace{0.5cm}}

\newcommand{\pathaveA}[1]{
   \left\langle\!\!\!\left\langle \:#1\:
   \right\rangle\!\!\!\right\rangle_{t} }

\newcommand{\vspfigA}{\vspace{0cm}}  
\newcommand{\vspfigB}{\vspace{0.1cm}} 
\newcommand{\vspfigC}{\vspace{0.2cm}}
\newcommand{\widthfigA}{0.48\textwidth}
\newcommand{\widthfigB}{0.4\textwidth}
\newcommand{\widthtableA}{2.3cm} 
\newcommand{\widthtableB}{4.0cm} 
\newcommand{\widthtableC}{2.0cm} 
\newcommand{\widthtableD}{4.5cm}

\begin{document}

%\baselineskip 0.8cm
%\baselineskip 0.46cm
%%%%%%%%%%%%%%%%%%%%%%%%%%%%%%%%%%%%%%%%%%%%%%%%%%%%%%%%%%%%%%%%%%%%%%

\title{ Inertial effects in nonequilibrium work fluctuations  
by a path integral approach }
\author{ Tooru Taniguchi and E. G. D. Cohen}
\affiliation{The Rockefeller University, 1230 York 
Avenue, New York, NY 10021, USA.} 
\date{\today}
\begin{abstract}
   Inertial effects in fluctuations of the work to sustain 
a system in a nonequilibrium steady state are discussed 
for a dragged massive Brownian particle model 
using a path integral approach. 
   We calculate the work distribution function 
in the laboratory and comoving frames 
and prove the asymptotic fluctuation theorem for these works  
for any initial condition. 
   Important and observable differences between 
the work fluctuations in the two frames appear  for finite times 
and are discussed concretely for  
a nonequilibrium steady state initial condition. 
   We also show that for finite times a 
time oscillatory behavior appears in the work distribution 
function for masses larger than a nonzero critical value.   
\end{abstract}

\pacs{ 
05.70.Ln, %Nonequilibrium and irreversible thermodynamics 
05.40.-a, %Fluctuation phenomena, random processes, noise, 
          %and Brownian motion 
05.10.Gg  %Stochastic analysis methods (Fokker-Planck, Langevin, etc.)
%02.50.-r %Probability theory, stochastic processes, and statistics 
%05.70.-a %Thermodynamics 
%05.40.Jc %Brownian motion 
%
%\vspace{0.1cm} 
%Keywords: 
%inertial effects;
%critical mass;
%nonequilibrium work fluctuations and theorem;
%%asymptotic fluctuation theorem; 
%path integration; 
%laboratory and comoving frames
%%nonequilibrium Onsager-Machlup theory; 
%%nonequilibrium steady state thermodynamics; 
%%nonequilibrium detailed balance;  
}

\vspace{1cm}
\maketitle
%%%%%%%%%%%%%%%%%%%%%%%%%%%%%%%%%%%%%%%%%%%%%%%%%%%%%%%%%%%%%%%%%%%%%%
\section{Introduction}

   In recent years, fluctuations in nonequilibrium 
systems have drawn considerable attention 
to a new kind of fluctuation theorems. 
   These fluctuation theorems are asymmetric relations 
for the distribution functions for work, heat, etc., and   
may be satisfied even far from equilibrium states 
or for small systems in which the magnitude 
of the fluctuations can be large. 
   These fluctuation theorems have been proved 
for deterministic thermostated systems \cite{ECM93,ES94,GC95} 
as well as for stochastic systems \cite{K98,LS99}, 
and have also been discussed in  
connection with the Onsager-Machlup fluctuation theory \cite{TC07a}. 
   Moreover, experimental confirmations for these theorems 
have been obtained \cite{CL98,WSM02,ST05,SST05}. 
   It has also been shown that the fluctuation theorems include  
the fluctuation-dissipation theorem, as well as 
Onsager's reciprocal relations, near equilibrium states 
\cite{ECM93,LS99,G96}.  
 
   In our previous paper \cite{TC07a}, 
based on a generalization of the Onsager-Machlup theory 
for fluctuations around equilibrium 
to those around nonequilibrium steady states 
using a path integral approach, 
we discussed fluctuation theorems for a stochastic dynamics 
described by a Langevin equation. 
    For a Brownian particle driven by a mechanical force 
$F(x_{s},s)$, the Langevin equation for the particle 
position $x_{s}$ at time $s$ is of the general form  
\begin{eqnarray}
   m\frac{d^{2} x_{s}}{ds^{2}} 
   = - \alpha \frac{d x_{s}}{ds} + F(x_{s},s) + \zeta_{s} 
\label{LangeEquat0a}
\end{eqnarray}
with the mass $m$ of the particle, the friction coefficient 
$\alpha$ and a random noise $\zeta_{s}$. 
   In our previous paper, as a nonequilibrium model  
we considered a dragged Brownian particle,  
in which the mechanical force is given by a harmonic force 
$F(x_{s},s)=-\kappa (x_{s}-vs)$  
with the spring constant $\kappa$ and  
the dragging velocity $v$. 
   Furthermore we mainly considered 
this model under the over-damped assumption. 
   This assumption can be used for a dynamics 
on a much longer time scale than 
the inertial characteristic time   
$\tau_{m}\equiv m/\alpha$, and  the dynamical equation 
under this assumption is simply 
given by neglecting the inertial term containing the mass 
in Eq. (\ref{LangeEquat0a}), i.e. by    
\begin{eqnarray}
   \frac{d x_{s}}{ds} = \frac{1}{\alpha} F(x_{s},s) 
   + \frac{1}{\alpha} \zeta_{s} .
\label{LangeEquat0b}
\end{eqnarray}
   Equation (\ref{LangeEquat0b}) is much simpler 
than Eq. (\ref{LangeEquat0a}), 
but information of the system 
on the shorter time scale than $\tau_{m}$ 
is lost in Eq. (\ref{LangeEquat0b}). 
   It may be noted that Machlup and Onsager already 
developed their fluctuation theory around equilibrium
not only for the case corresponding to the over-damped case  
\cite{OM53} but also for the inertial case \cite{MO53}. 
   In our previous paper we discussed also    
a generalization of the Onsager-Machlup theory 
for nonequilibrium steady states including   
the inertial term \cite{TC07a}. 
   However, there we treated only one type of  
fluctuation theorem, the so called transient 
fluctuation theorem \cite{ES94}, which is restricted to  
equilibrium initial conditions. 
   Another fluctuation theorem, 
the asymptotic fluctuation theorem \cite{GC95}, which holds for 
any initial condition (including a 
nonequilibrium steady state%
\footnote{
   A fluctuation theorem for a nonequilibrium steady 
state initial condition has been called 
the steady state fluctuation theorem 
(or the Gallavotti-Cohen fluctuation theorem \cite{GC95}), 
which is a special case of asymptotic fluctuation theorems.  
}
$\!$), was not discussed   
for inertial cases in Ref. \cite{TC07a}. 
   Different from the transient fluctuation theorem, 
which is correct for all times as a mathematical identity 
\cite{CG99}, the asymptotic fluctuation theorem is satisfied 
in the long time limit only. 
   However, 
as we will discuss in this paper, 
a variety of interesting inertial effects appear for finite times 
for a nonequilibrium initial condition, 
before the asymptotic fluctuation theorem is achieved.  
   Although there are some results for fluctuation theorems  
for stochastic systems including inertia \cite{ZBC05,DJG06}, 
the asymptotic fluctuation theorem with inertia has not been 
discussed fully in connection with the Onsager-Machlup theory 
so far. 

   The purpose of this paper is therefore to 
discuss, in the context of the Onsager-Machlup 
path integral approach, inertial effects 
in nonequilibrium steady state work fluctuations, 
including the asymptotic fluctuation theorem. 
   For these discussions we use the Langevin equation 
(\ref{LangeEquat0a}) for a dragged Brownian particle 
without the over-damped assumption. 
   The work distribution function is calculated 
explicitly for any initial condition, 
and its finite time properties are investigated. 
   As an important inertial effect  
we show a critical value of mass above which 
the work distribution function shows a time-oscillatory behavior.

   The nonequilibrium work used in this paper is based on 
the generalized Onsager-Machlup theory,  
as obtained in our previous paper \cite{TC07a}. 
   In that paper we considered two kinds of work 
in two different frames: 
(A) the work $\mathcal{W}_{l}$ done in the laboratory frame ($l$)
%with a nonzero average velocity of particle 
%in a noneqilibrium steady state,  
and (B) the work $\mathcal{W}_{c}$ done 
in the comoving frame ($c$) where the average velocity of 
the Brownian particle 
is zero in a nonequilibrium steady state. 
   A difference between these two works is that 
$\mathcal{W}_{c}$ includes a d'Alembert-like force,  
which is absent in $\mathcal{W}_{l}$. 
   In this paper, we show that both the works $\mathcal{W}_{l}$ 
and $\mathcal{W}_{c}$ satisfy the asymptotic fluctuation theorem. 
   We also discuss dramatic differences  
between the work distribution functions for $\mathcal{W}_{l}$ 
and $\mathcal{W}_{c}$ 
for finite times. 
 
   The outline of this paper is as follows. 
   In Sec. \ref{DraggBrown} we introduce a dragged Brownian 
particle model with inertia,  
and treat its dynamics using a path integral.   
   In Sec. \ref{WorkInert} we introduce the works done  
in the laboratory and comoving frames
and calculate their distribution functions. 
   In Sec. \ref{AsympFluct} we prove 
the asymptotic work fluctuation theorem. 
   In Sec. \ref{InertEffec} we discuss inertial effects 
in the work distribution functions for finite times. 
   Finally, Sec. \ref{ConclRemar} is devoted to 
a summary and some remarks on this paper.

%%%%%%%%%%%%%%%%%%%%%%%%%%%%%%%%%%%%%%%%%%%%%%%%%%%%%%%%%%%%%%%%%%%%%%
\section{Dragged Brownian Particle with Inertia} 
\label{DraggBrown}

   We consider a Brownian particle confined by 
a harmonic potential, which moves  
with a constant velocity $v$ through a fluid, 
as discussed in our previous paper \cite{TC07a}. 
   The dynamics of this particle is described by 
a Langevin equation 
\begin{eqnarray}
   m\frac{d^{2} x_{s}}{ds^{2}} 
   = - \alpha \frac{d x_{s}}{ds} 
     - \kappa \left(x_{s}-vs\right) + \zeta_{s} .
\label{LangeEquat1}
\end{eqnarray}
%
%for the particle position $x_{s}$ at time $s$, 
%by Eq. (\ref{LangeEquat0a}) with a harmonic 
%force $F(x_{s},s)=-\kappa (x_{s}-vs)$. 
%where $m$ is the mass of the particle, $\alpha$ is the friction 
%constant, and $\kappa$ is the spring constant of 
%the harmonic potential. 
   Here, we assume that $\zeta_{s}$ is 
the Gaussian-white random force whose probability functional 
$P_{\zeta}(\{\zeta_{s}\})$ 
for $\{\zeta_{s}\}_{s\in[t_{0},t]}$ is given by 
\begin{eqnarray}
   P_{\zeta}(\{\zeta_{s}\}) 
   = C_{\zeta} \exp\left( -\frac{\beta}{4\alpha}\int_{t_{0}}^{t}ds\;
   \zeta_{s}^{2} \right)
\label{NoiseProba1}
\end{eqnarray}
with the normalization coefficient $C_{\zeta}$ 
and  the inverse temperature $\beta \equiv 1/(k_{B}T)$,  
where $k_{B}$ is the Boltzmann's constant 
and $T$ is the temperature of the heat reservoir. 
   [Note that the coefficient $C_{\zeta}$ can depend on 
the initial time $t_{0}$ and the final time $t$, 
but such time dependences in $C_{\zeta}$, 
as well as in similar coefficients $C_{x}$ and $C_{\mathcal{E}}$ 
introduced later, are suppressed.]
   It follows from Eq. (\ref{NoiseProba1}) that the first two 
auto-correlation functions of the random force $\zeta_{s}$ 
are given by   
$\langle \zeta_{s} \rangle = 0 $ and 
$\langle \zeta_{s_{1}}\zeta_{s_{2}} \rangle 
= (2\alpha/\beta) \delta(s_{1}-s_{2})$
with the notation $\langle \cdots\rangle$ 
for an initial ensemble average. 
%   The left-hand side of Eq. (\ref{LangeEquat1}) 
%expresses the effect of inertia for the Brownian particle 
%and the case neglecting such a term is the over-damped 
%case, treated in Ref. \cite{TC07a}. 

   Now, we consider the probability functional $P_{x}(\{x_{s}\})$ 
for a path $\{x_{s}\}_{s\in[t_{0},t]}$ of the particle position 
$x_{s}$. 
   By inserting Eq. (\ref{LangeEquat1})  
into Eq. (\ref{NoiseProba1}) and interpreting 
the probability functional $P_{\zeta}(\{\zeta_{s}\})$ 
for $\zeta_{s}$ as the probability functional 
$P_{x}(\{x_{s}\})$ for $x_{s}$, we obtain, 
apart from a normalization coefficient,  
\begin{eqnarray}
   && P_{x}(\{x_{s}\}) 
   \nonumber \\
   &&  = C_{x} \exp\left[ -\frac{1}{4D}\int_{t_{0}}^{t}ds\; 
      \left(\dot{x}_{s} + \frac{x_{s}-vs}{\tau_{r}}
      + \frac{m}{\alpha}\ddot{x}_{s} 
      \right)^{2}\right] \spaEq
\label{PathProba1}
\end{eqnarray}
with $\dot{x}_{s}\equiv dx_{s}/ds$,  
$\ddot{x}_{s}\equiv d^{2}x_{s}/ds^{2}$ 
and the normalization coefficient $C_{x}$. 
   Here, $D\equiv k_{B}T/\alpha$ is the diffusion constant 
given by the Einstein relation 
and $\tau_{r}\equiv\alpha/\kappa$ is the relaxation time 
in the over-damped case. 
   For another derivation of Eq. (\ref{PathProba1}) 
via a Fokker-Planck equation corresponding to the 
Langevin equation, see, for example, Ref. \cite{R89}.  

   For systems whose dynamics is expressed by 
a second-order Langevin equation,  
like Eq. (\ref{LangeEquat1}), we introduce 
the path integration of any functional $X(\{x_{s}\})$ as 
%
%\begin{eqnarray}
$
   \int_{(x_{t_{0}},\dot{x}_{t_{0}})=(x_{i},p_{i}/m)}^{
   (x_{t},\dot{x}_{t})=(x_{f},p_{f}/m)}
   \mathcal{D}x_{s} \;$ $X(\{x_{s}\})
$, 
%   \nonumber
%%\label{}
%\end{eqnarray}
% 
with respect to paths $\{x_{s}\}_{s\in(t_{0},t)}$ 
satisfying the initial ($i$) condition 
$(x_{t_{0}},\dot{x}_{t_{0}})=(x_{i},p_{i}/m)$ 
and the final ($f$) condition  
$(x_{t},\dot{x}_{t})=(x_{f},p_{f}/m)$. 
%   (Here, note that we must specify the positions 
%and velocities of the particle at the initial time $t_{0}$ 
%and the final time $t$ in order to define a path 
%in the case with inertia.)
   Using this notation for the functional integral, 
the functional average $\pathaveA{X(\{x_{s}\})}$  
over all possible paths $\{x_{s}\}_{s\in(t_{0},t)}$,  
as well as averages over the initial and final 
positions and momenta of the particle  
is represented by 
\begin{eqnarray}
   \pathaveA{X(\{x_{s}\})} &\equiv& 
      \int\!\int dx_{i}dp_{i} 
      \int_{(x_{t_{0}},\dot{x}_{t_{0}})=(x_{i},p_{i}/m)}^{
      (x_{t},\dot{x}_{t})=(x_{f},p_{f}/m)}\; 
      \mathcal{D}x_{s} 
      \nonumber \\
   &&\spaEq\times 
      \int\!\int dx_{f}dp_{f} \; X(\{x_{s}\})
      \nonumber \\
   &&\spaEq\times
       P_{x}(\{x_{s}\}) f(x_{i},p_{i},t_{0})
\label{FunctAvera1}
\end{eqnarray}
with the initial distribution function $f(x_{i},p_{i},t_{0})$ 
for the particle position $x_{i}$ and momentum $p_{i}$. 
   The normalization condition  
to specify the coefficient $C_{x}$ 
of the distribution functional (\ref{PathProba1})  
is given by $\pathaveA{1}=1$ using the 
notation (\ref{FunctAvera1}) as well as the normalization condition 
$\int\!\int dx_{i}dp_{i}\; $ $f(x_{i},p_{i},t_{0})=1$ 
for the initial distribution function $f(x_{i},p_{i},t_{0})$. 

   This finishes the introduction of our model and its dynamics. 
   In the next section \ref{WorkInert} 
we introduce the work done on this system  
and calculate its probability distribution.

%%%%%%%%%%%%%%%%%%%%%%%%%%%%%%%%%%%%%%%%%%%%%%%%%%%%%%%%%%%%%%%%%%%%%%
\section{Work Distribution} 
\label{WorkInert}
\subsection{Work to Drag a Brownian Particle and its Distribution} 
\label{WorkDragParti}

   In our previous paper \cite{TC07a}, we considered 
the work $\mathcal{W}$ to move the confining potential 
with a velocity $v$ in two frames; 
the laboratory frame using the particle 
position $x_{s}$ and the 
comoving frame using the particle position 
$y_{s}\equiv x_{s}-vs$ at time $s$. 
   Based on a generalized Onsager-Machlup theory, 
we showed in Ref. \cite{TC07a} that the work $\mathcal{W}_{l}$ 
done in the laboratory frame is 
given by $\int_{t_{0}}^{t}ds\; [-\kappa (x_{s}-vs)] v$,  
and the work $\mathcal{W}_{c}$ done in the comoving frame  
is given by $\int_{t_{0}}^{t}ds\; (-\kappa y_{s}-m\ddot{y}_{s}) v$ 
with $\ddot{y}_{s}\equiv d^{2} y_{s}/ds^{2} = \ddot{x}_{s}$, 
leading to a difference between the work $\mathcal{W}$ 
in these two frames by an inertial or d'Alembert-like force 
$-m\ddot{y}_{s}$ . 
%   Here, the term $-m \ddot{x}_{s}$ is the d'Alembert force 
%as an inertial force, which also appears in Onsager and 
%Machlup's original work \cite{MO53}. 
   To understand this difference in a concise way, note first that 
by the energy conservation law, the work $\mathcal{W}$ is given by 
the heat $Q$ and the energy difference $\Delta E$, namely 
by $\mathcal{W} = Q+\Delta E$, where the energy difference $\Delta E$ 
is the sum of the kinetic energy difference $\Delta K$ and 
the potential energy difference $\Delta U$, i.e. 
$\Delta E = \Delta U + \Delta K$. 
   Here, the kinetic energy difference $\Delta K 
= \Delta K_{c}$ and $\Delta K_{l}$ 
in the comoving frame 
and the laboratory frame 
are given by $(m \dot{y}_{t}^{2}/2) 
- (m \dot{y}_{t_{0}}^{2}/2)$ 
and $(m \dot{x}_{t}^{2}/2) - (m \dot{x}_{t_{0}}^{2}/2)$, 
respectively,  
%\footnote{
%   The particle velocity  
%$\dot{y}_{s} (= \dot{x}_{s}-v)$ obtained by subtraction 
%of the average velocity $v$ in the comoving frame 
%is called the ``thermal'' (or ``peculiar'') velocity,  
%and the corresponding kinetic energy 
%$m \dot{y}_{s}^{2}/2$  is used to define the so-called 
%kinetic temperature \cite{Eva90a} in nonequilibrium statistical 
%thermodynamics.  
%}
 so that we obtain the relation  
\begin{eqnarray}
   \Delta K_{c} = \Delta K_{l} 
   -\int_{t_{0}}^{t} ds\; m\ddot{x}_{s} v .  
\label{KinetDiffe1}
\end{eqnarray}
   Equation (\ref{KinetDiffe1}) means that the kinetic energy 
difference $\Delta K$ depends on the frames and 
its frame-difference is determined by the d'Alembert-like force 
$-m \ddot{x}_{s}$ as a purely inertial effect. 
   This frame-difference of $\Delta K$ also appears 
in the work, and leads to  
the relation $\mathcal{W}_{c} = \mathcal{W}_{l} 
-\int_{t_{0}}^{t} ds\; m\ddot{x}_{s} v$. 
% between the two kinds of work in the different frames. 
%   It should be emphasized that this difference 
%of the works in the different frames appears 
%as a purely inertial effect, and that the two works 
%$\mathcal{W}_{c}$ and $\mathcal{W}_{l}$ coincide 
%with each other in the over-damped case.  
   A more complete explanation for this frame-dependence 
of the work is given in Ref. \cite{TC07a}, 
based on a nonequilibrium generalization of the 
detailed balance condition.  
   
   To discuss these two different kinds of work 
done in the laboratory and comoving frames simultaneously  
in this paper, we consider the work defined in general by 
\begin{eqnarray}
   \mathcal{W}(\{x_{s}\}) = \int_{t_{0}}^{t}ds\;
   \left[-\kappa (x_{s}-vs) 
   -(1-\vartheta) m\ddot{x}_{s}\right] v ,
\label{Work1}
\end{eqnarray}
which gives the work $\mathcal{W}_{l}$ done in the laboratory case 
$(\vartheta=1)$ as well as the work $\mathcal{W}_{c}$ 
done in the comoving case 
$(\vartheta=0)$ by changing value of the parameter $\vartheta$.% 
\footnote{
   The parameter $\vartheta$ in Eq. (\ref{Work1}) is chosen 
   in a way consistent to that in our previous paper \cite{TC07a}.  
} 

   Using the functional average defined by Eq. (\ref{FunctAvera1}), 
the probability distribution $P_{w}(W)$ for the dimensionless work 
$\beta\mathcal{W}(\{x_{s}\})$ is given by 
\begin{eqnarray}
   P_{w}(W,t) = \pathaveA{\delta\!\left(
   W-\beta\mathcal{W}(\{x_{s}\})\right)} .
\label{DistriWork1}
\end{eqnarray}
   For later calculative convenience, 
we introduce a Fourier transformation 
$\mathcal{E}_{w}(i\lambda,t)$ of the work distribution function  
$P_{w}(W,t)$  through the function 
$\mathcal{E}_{w}(\lambda,t)$ defined by 
\begin{eqnarray}
   \mathcal{E}_{w}(\lambda,t) \equiv 
%   \pathaveA{\exp\left[-\lambda \beta \mathcal{W}(\{x_{s}\})
%   \right]} = 
   \pathaveA{e^{-\lambda \beta \mathcal{W}(\{x_{s}\})}} , 
\label{EFunctWork1}
\end{eqnarray}
so that 
the work distribution function $P_{w}(W)$ can be represented as
\begin{eqnarray}
   P_{w}(W,t) 
   =\frac{1}{2\pi}\int_{-\infty}^{+\infty} d\lambda\; 
      \mathcal{E}_{w}(i\lambda,t) \; e^{i\lambda W} .
      %\exp(i\lambda W) .
\label{DistriWork2}
\end{eqnarray}
   The function $\mathcal{E}_{w}(\lambda,t)$ 
can be also regarded as a generating function for the work 
$\mathcal{W}(\{x_{s}\})$. 
   By Eq. (\ref{EFunctWork1}) we obtain a useful identity 
\begin{eqnarray}
  \mathcal{E}_{w}(0,t) = 1 
\label{IdentE1}
\end{eqnarray}
used to determine a normalization constant 
later [Eq. (\ref{NormaConst1})]. 
%   In the next section we carry out a functional integral 
%involving in the functional average in Eq. (\ref{EFunctWork1}), 
%and then calculate the work distribution function 
%by Eq. (\ref{DistriWork2}). 

%%%%%%%%%%%%%%%%%%%%%%%%%%%%%%%%%%%%%%%%%%%%%%%%%%%%%%%%%%%%%%%%%%%%%%
\subsection{Path Integral Analysis for Work Distribution} 

   To calculate the function 
$\mathcal{E}_{w}(\lambda,t)$ from Eq. (\ref{EFunctWork1}), 
we first note that 
\begin{eqnarray}
    \mathcal{E}_{w}(\lambda,t) &=& 
      C_{x}\int\!\int dx_{i}dp_{i} 
      \int_{(x_{t_{0}},\dot{x}_{t_{0}})=(x_{i},p_{i}/m)}^{
      (x_{t},\dot{x}_{t})=(x_{f},p_{f}/m)}\; 
      \mathcal{D}x_{s} 
      \nonumber \\
   &&\spaEq\times \int\!\int dx_{f}dp_{f} \; f(x_{i},p_{i},t_{0}) 
      \nonumber \\
   &&\spaEq\times
     \exp\left[
     \int_{t_{0}}^{t}ds\; L(\ddot{x}_{s},\dot{x}_{s},x_{s},s)\right]
\label{EFunctWork2}
\end{eqnarray}
by Eqs. (\ref{PathProba1}), (\ref{FunctAvera1}), (\ref{Work1})  
and (\ref{EFunctWork1}). 
   Here, $ L(\ddot{x}_{s},\dot{x}_{s},x_{s},s)$ is defined by 
\begin{eqnarray}
    L(\ddot{x}_{s},\dot{x}_{s},x_{s},s) &\equiv&
      -\frac{1}{4D}\left(\dot{x}_{s} + \frac{x_{s}-v s}{\tau_{r}}
      + \frac{m}{\alpha}\ddot{x}_{s} \right)^{2} 
      \nonumber \\
   &&\spaEq
      +\lambda \beta \left[\kappa (x_{s}-v s) 
      +(1-\vartheta) m\ddot{x}_{s}\right] v , 
      \nonumber \\
\label{LagraFunct1}
\end{eqnarray}
which may be interpreted as a Lagrangian function including 
a Lagrange multiplier $\lambda$ 
due to the restriction of the delta function for work 
in Eq. (\ref{DistriWork1}) \cite{TC07a}%
\footnote{
   In Ref. \cite{TC07a} we called only the first term on 
the right-hand side of Eq. (\ref{LagraFunct1}) 
the Lagrangian function in the Onsager-Machlup theory, 
which is directly connected to a transition probability. 
}. 
   Here, as elsewhere in this paper, 
the dependence of $L(\ddot{x}_{s},\dot{x}_{s},x_{s},s)$ 
on the parameters $v$, $\vartheta$, etc., has not been explicitly indicated on the  left-hand side of Eq. (\ref{LagraFunct1}). 

   The first step to calculate the function 
$\mathcal{E}_{w}(\lambda,t)$  
is to specify the most-contributing path 
$\{x_{s}^{*}\}_{s\in[t_{0},t]}$ in the path integral 
involved on the right-hand side of Eq. (\ref{EFunctWork2}).  
   Such a special path $\{x_{s}^{*}\}_{s\in[t_{0},t]}$ 
is introduced as the one satisfying the variational principle
\begin{eqnarray}
   \delta \int_{t_{0}}^{t}ds\; 
   L(\ddot{x}_{s}^{*},\dot{x}_{s}^{*},x_{s}^{*},s) = 0 
\label{MinimActio1}
\end{eqnarray}
with the four boundary conditions 
$x_{t_{0}}^{*}=x_{i}$, $\dot{x}_{t_{0}}^{*}=p_{i}/m$, 
$x_{t}^{*}=x_{f}$ and $\dot{x}_{t}^{*}=p_{f}/m$. 
%$x_{t_{0}}^{*}=x_{i}$, $\dot{x}_{t_{0}}^{*}=p_{i}/m$, 
%$x_{t}^{*}=x_{f}$ and $\dot{x}_{t}^{*}=p_{f}/m$. 
   In a way similar to derive the Euler-Lagrange equation 
from the minimum action principle in analytical 
mechanics \cite{LL69}, Eq. (\ref{MinimActio1}) leads to 
\begin{eqnarray}
   &&\frac{d^{2}}{ds^{2}}
      \frac{\partial L(\ddot{x}_{s}^{*},\dot{x}_{s}^{*}
      ,x_{s}^{*},s) }{\partial \ddot{x}_{s}^{*}} 
      - \frac{d}{ds}\frac{\partial L(\ddot{x}_{s}^{*},\dot{x}_{s}^{*}
      ,x_{s}^{*},s) }{\partial \dot{x}_{s}^{*}} 
      \nonumber \\
   &&\spaEq + \frac{\partial L(\ddot{x}_{s}^{*},\dot{x}_{s}^{*}
      ,x_{s}^{*},s) }{\partial x_{s}^{*}} 
   = 0 
\label{LagraEquat1}
\end{eqnarray}
for the Lagrangian function (\ref{LagraFunct1}).  
   Inserting Eq. (\ref{LagraFunct1}) into Eq. 
(\ref{LagraEquat1}) we obtain a 
fourth-order linear differential equation 
\begin{eqnarray}
    \tau_{m}^{2}\frac{d^{4}\tilde{x}_{s}^{*}}{ds^{4}}
    - \left(1-2\frac{\tau_{m}}{\tau_{r}} \right)
    \frac{d^{2}\tilde{x}_{s}^{*}}{ds^{2}} 
    + \frac{1}{\tau_{r}^{2}} \tilde{x}_{s}^{*} 
    =0
\label{LagraEquat5}
\end{eqnarray}
for the function $\tilde{x}_{s}^{*}$ of $s$,  
which is defined by 
\begin{eqnarray}
   \tilde{x}_{s}^{*} \equiv x_{s}^{*} - v s + (1-2\lambda) v\tau_{r} ,  
\label{TildeFunctX1}
\end{eqnarray}
using the inertial characteristic time $\tau_{m}\equiv m/\alpha$.   

   We consider solutions of Eq. (\ref{LagraEquat5}) 
of the form $\exp(\nu s)$. 
   Inserting $\tilde{x}_{s}^{*}=\exp(\nu s)$ 
into Eq. (\ref{LagraEquat5}) we obtain the quadratic equation 
\begin{eqnarray}
   &&\tau_{m}^{2}\nu^{4} 
   - \left(1-2\frac{\tau_{m}}{\tau_{r}} \right)\nu^{2}
   + \frac{1}{\tau_{r}^{2}} 
   \nonumber \\
   &&\spaEq = \left(\tau_{m} \nu^{2} + \nu +\frac{1}{\tau_{r}}\right)
   \left(\tau_{m} \nu^{2} - \nu +\frac{1}{\tau_{r}}\right) 
   \nonumber \\
   &&\spaEq 
   = 0
\label{QuartEquat1}
\end{eqnarray}
for $\nu$. 
   The solutions of Eq. (\ref{QuartEquat1}) are 
$\nu = \nus{+}, \nus{-}, -\nus{-}, -\nus{+}$ 
using $\nus{\pm}$ defined by 
\begin{eqnarray}
   \nus{\pm} = \frac{1}{2\tau_{m}}
   \left(1\pm\sqrt{1-4\frac{\tau_{m}}{\tau_{r}}}\;\right) .
\label{CoeffNuPM}  
\end{eqnarray} 
   The general solution of the fourth-order 
differential equation (\ref{LagraEquat5}) is 
represented as a superposition of 
these special solutions $\exp(\nu s)$, 
$\nu = \nus{+}, \nus{-}, -\nus{-}, -\nus{+}$, namely  
\begin{eqnarray}
   \tilde{x}_{s}^{*} = C_{1}e^{\nus{+}s}  
   +C_{2} e^{\nus{-}s} + C_{3}e^{-\nus{-}s} 
   +C_{4} e^{-\nus{+}s} 
\label{SolutTildeX1}
\end{eqnarray}
with constants $C_{j}$, $j=1,2,3,4$. 
   Using Eqs. (\ref{TildeFunctX1}) and  
(\ref{SolutTildeX1}) and introducing 
the four dimensional vector 
$\mathbf{C} \equiv (C_{1}\; C_{2}\; C_{3}\; C_{4})^{T}$,%
\footnote{
   In this paper, $X^{T}$ means the transposed matrix (or vector) 
of any matrix (or vector) $X$.
} 
we can rewrite  
\begin{eqnarray}
   x_{s}^{*} =  \mathbf{C}^{T}\mathbf{K}_{s} 
   + v s - (1-2\lambda) v\tau_{r} 
\label{FunctX1}
\end{eqnarray}
where the vector $\mathbf{K}_{s}$ is defined by 
\begin{eqnarray}
   \mathbf{K}_{s} &\equiv& 
      \left(\begin{array}{c}
         e^{\nus{+}s}  \\ e^{\nus{-}s} \\ 
         e^{-\nus{-}s} \\ e^{-\nus{+}s}
   \end{array}\right) . 
   \label{VectoK1}
\end{eqnarray}
   The constant vector $\mathbf{C}$
is determined by the four boundary conditions 
for $x_{s}^{*}$ and we obtain 
\begin{eqnarray}
   \mathbf{C} &=& A_{t}^{-1} \mathbf{B}_{if}^{(1-2\lambda)}
      \label{ConstC1}
\end{eqnarray}
where the matrix $A_{t}$ is defined by 
\begin{eqnarray}
   A_{t}&\equiv& 
      \left(\begin{array}{cccc}
      e^{\nus{+}t_{0}} & e^{\nus{-}t_{0}} & 
         e^{-\nus{-}t_{0}} & e^{-\nus{+}t_{0}} \\ 
      \nus{+}e^{\nus{+}t_{0}} & 
         \nus{-}e^{\nus{-}t_{0}} & 
         -\nus{-}e^{-\nus{-}t_{0}} & 
         -\nus{+}e^{-\nus{+}t_{0}} \\ 
      e^{\nus{+}t} & e^{\nus{-}t} & 
         e^{-\nus{-}t} & e^{-\nus{+}t} \\ 
      \nus{+}e^{\nus{+}t} & 
         \nus{-}e^{\nus{-}t} & 
         -\nus{-}e^{-\nus{-}t} & 
         -\nus{+}e^{-\nus{+}t} 
   \end{array}\right) \spaEq
\label{MatriA1} 
\end{eqnarray}
and the vector $\mathbf{B}_{if}^{(z)}$ is defined by 
\begin{eqnarray}
   \mathbf{B}_{if}^{(z)}&\equiv& 
      \left(\begin{array}{c}
         x_{i} -vt_{0}\\ p_{i}/m -v\\ x_{f} -vt\\ p_{f}/m -v
      \end{array}\right) 
   +z v\tau_{r}
      \left(\begin{array}{c}
         1\\ 0\\1\\ 0
      \end{array}\right) .
\label{VectoBJ}
\end{eqnarray}
   It may be noted that the first component $x_{i} -vt_{0}$ 
and the second component $(p_{i}/m) -v$ 
(the third component $x_{f} -vt_{0}$ 
and the fourth component $(p_{f}/m) -v$ ) of the vector 
$\mathbf{B}_{if}^{(0)}$ can be regarded as the initial 
(final) position 
and velocity of the particle in the comoving frame, respectively. 

   As the next step, we represent a path 
$\{x_{s}\}_{s\in [t_{0},t]}$ as the sum of the most contributing 
path $\{x_{s}^{*}\}_{s\in [t_{0},t]}$ given by Eq. (\ref{FunctX1}) 
and its deviation $\{\Delta x_{s}\}_{s\in [t_{0},t]}$ defined by  
\begin{eqnarray}
   \Delta x_{s} \equiv x_{s} - x_{s}^{*} , 
\label{VariaDeltaX1}
\end{eqnarray}
where the variable $\Delta x_{s}$ 
satisfies the four boundary conditions 
$\Delta x_{t_{0}} = \Delta x_{t} = 0 $ and 
$\Delta\dot{x}_{t_{0}} = \Delta\dot{x}_{t} = 0$
with $\Delta\dot{x}_{s}\equiv d \Delta x_{s}/ds$. 
   Using this variable  $\Delta x_{s}$, 
the complete time integral $\int_{t_{0}}^{t}ds\; 
L(\ddot{x}_{s},\dot{x}_{s},x_{s},s)$ of the Lagrangian function 
can be represented as   
\begin{eqnarray}
   &&\int_{t_{0}}^{t}ds\;L(\ddot{x}_{s},\dot{x}_{s},x_{s},s) 
   \nonumber \\
   &&\spaEq 
      = \int_{t_{0}}^{t}ds\; 
      L(\ddot{x}_{s}^{*},\dot{x}_{s}^{*},x_{s}^{*},s)
      \nonumber \\
   && \spaEq\spaEq
      -\frac{1}{4D}\int_{t_{0}}^{t}ds\; \left(
      \Delta\dot{x}_{s}
      +\frac{1}{\tau_{r}}\Delta x_{s}
      + \frac{m}{\alpha}\Delta\ddot{x}_{s}
      \right)^{2} .
\label{LagraFunct2}
\end{eqnarray}
in terms of the two variables $x_{s}^{*}$ and $\Delta x_{s}^{*}$.
   Inserting Eq. (\ref{LagraFunct2}) 
into Eq. (\ref{EFunctWork2}) we obtain 
\begin{eqnarray}
    \mathcal{E}_{w}(\lambda,t) &=& 
      C_{\mathcal{E}}\int\!\int dx_{i}dp_{i} \int\!\int dx_{f}dp_{f} 
      \nonumber \\
   &&\spaEq\times
     f(x_{i},p_{i},t_{0}) \exp\left[
     \int_{t_{0}}^{t}ds\; 
     L(\ddot{x}_{s}^{*},\dot{x}_{s}^{*},x_{s}^{*},s)\right]
     \nonumber \\
\label{EFunctWork3}
\end{eqnarray}
where $C_{\mathcal{E}}$ is defined by 
%%
%\begin{eqnarray}
$   C_{\mathcal{E}} \equiv C_{x}
      \int_{(\Delta x_{t_{0}},\Delta\dot{x}_{t_{0}})=(0,0)}^{
      (\Delta x_{t},\Delta\dot{x}_{t})=(0,0)} 
      \mathcal{D}\Delta x_{s}\;   
%      \nonumber \\
%   &&\spaEq\times
$ $
      \exp[ - (1/4D)
     \int_{t_{0}}^{t}ds$ $\int_{t_{0}}^{t}ds\; 
      (\Delta\dot{x}_{s}
      +(1/\tau_{r})\Delta x_{s}
      + (m/\alpha)\Delta\ddot{x}_{s})^{2}] 
$
%\label{TildeCx1}
%\end{eqnarray}
%
and is independent of $\lambda$.  
   In the expression (\ref{EFunctWork3}), 
the contributions of 
the deviations $\Delta x_{s}$ to  
the path integral in the function 
$\mathcal{E}_{w}(\lambda,t)$ are included  
only in the coefficient $C_{\mathcal{E}}$. 

   Next, we calculate the quantity 
$\int_{t_{0}}^{t}ds\;L(\ddot{x}_{s}^{*},\dot{x}_{s}^{*},x_{s}^{*},s)$ 
using Eq. (\ref{FunctX1}), and then the function 
$\mathcal{E}_{w}(\lambda,t)$ given by Eq. (\ref{EFunctWork3}). 
   For such a calculation, using Eq. (\ref{FunctX1}) 
we first note that  
\begin{eqnarray}
   \frac{d x_{s}^{*}}{ds} 
      &=& \mathbf{C}^{T}\Theta\mathbf{K}_{s} + v , 
      \label{DerivX1} \\
   \frac{d^{2}x_{s}^{*}}{ds^{2}} 
      &=& \mathbf{C}^{T}\Theta^{2}\mathbf{K}_{s} 
      \label{DerivX2}
\end{eqnarray}
where the $4\times 4$ matrix $\Theta$ 
is defined by 
\begin{eqnarray}
   \Theta \equiv 
      \left(\begin{array}{cccc}
         \nus{+} &0&0&0 \\ 
         0& \nus{-} &0&0 \\ 
         0&0& -\nus{-} &0 \\ 
         0&0&0& -\nus{+}
   \end{array}\right) .
\label{MatriTheta1}
\end{eqnarray}
   Then, using Eqs. (\ref{FunctX1}), 
(\ref{DerivX1}) and (\ref{DerivX2}) we obtain 
\begin{eqnarray}
   \dot{x}_{s}^{*} + \frac{x_{s}^{*}-vs}{\tau_{r}}
      + \frac{m}{\alpha}\ddot{x}_{s}^{*} 
      = \mathbf{C}^{T}\Gamma\mathbf{K}_{s} + 2\lambda v 
\label{LagraCalcu1}
\end{eqnarray}
where the matrix $\Gamma$ is introduced as  
\begin{eqnarray}
   \Gamma &\equiv& 
      \tau_{m}\Theta^{2}+\Theta+\frac{1}{\tau_{r}}\mathcal{I} 
      \nonumber \\
   &=& 2 \left(\begin{array}{cccc}
         \nus{+} & 0 &0&0  \\
         0 & \nus{-}&0&0 \\
         0 & 0 &0&0\\
         0 & 0 &0& 0
      \end{array}\right) 
\label{functGamma1}
\end{eqnarray}
with the relation $\tau_{m}\nus{\pm}^{2}-\nus{\pm}+\tau_{r}^{-1} = 0$ 
and $\mathcal{I}$ the $4\times 4$ identity matrix. 
    Using Eqs. (\ref{LagraFunct1}), (\ref{FunctX1}), 
(\ref{DerivX2}) and  (\ref{LagraCalcu1}) we obtain 
\begin{eqnarray}
   L(\ddot{x}_{s}^{*},\dot{x}_{s}^{*},x_{s}^{*},s)  
      &=& -\frac{1}{4}\alpha\beta\mathbf{C}^{T}
      \Gamma\mathbf{K}_{s} \mathbf{K}_{s}^{T} \Gamma\mathbf{C} 
   \nonumber \\
   &&\spaEq 
      - \alpha\beta\lambda v \mathbf{C}^{T}
      \left[\tau_{m}\vartheta\Theta^{2}+\Theta\right]\mathbf{K}_{s}
      \nonumber \\
   &&\spaEq 
      - \lambda(1-\lambda) \alpha\beta v^{2} .
\label{LagraFunct4}
\end{eqnarray}
   Noting Eq. (\ref{ConstC1}) and that  
\begin{eqnarray}
   \int_{t_{0}}^{t}ds\; \mathbf{C}^{T}\Theta\mathbf{K}_{s}  
      &=& \int_{t_{0}}^{t}ds\;\frac{dx_{s}^{*}}{ds} 
      - v(t-t_{0}) 
      \nonumber \\
   &=& x_{f}-x_{i} - v(t-t_{0}) ,
      \label{BoundCondi5}\\
   \int_{t_{0}}^{t}ds\;\mathbf{C}^{T}\Theta^{2}\mathbf{K}_{s}  
      &=& \int_{t_{0}}^{t}ds\;\frac{d^{2}x_{s}^{*}}{ds^{2}}
      = \frac{p_{f}-p_{i}}{m}
      \label{BoundCondi6}
\end{eqnarray}
by Eqs. (\ref{DerivX1}) and (\ref{DerivX2}), 
we further obtain 
\begin{eqnarray}
   &&\int_{t_{0}}^{t}ds\; 
      L(\ddot{x}_{s}^{*},\dot{x}_{s}^{*},x_{s}^{*},s) 
      \nonumber \\
   &&\spaEq 
      = -\frac{1}{4}\alpha\beta 
      \left[\mathbf{B}_{if}^{(1-2\lambda)}\right]^{T}
      \Lambda_{t} \mathbf{B}_{if}^{(1-2\lambda)}
      - \alpha\beta\lambda v 
      \bfeta^{T}\mathbf{B}_{if}^{(0)}
      \nonumber \\
   &&\spaEq\spaEq 
      - \lambda(1-\lambda) \alpha\beta v^{2} (t-t_{0})
\label{ActioFunct2}
\end{eqnarray}
where the $4\times 4$ matrix $\Lambda_{t}$ 
and the vector $\bfeta$ are defined by 
\begin{eqnarray}
    \Lambda_{t} &\equiv& \left(A_{t}^{-1}\right)^{T}
       \Gamma\Phi_{t}
       \Gamma A_{t}^{-1} ,
       \label{MatriLambd1} \\
%    &\equiv& \left(A_{t}^{-1}\right)^{T}
%       \left(\tau_{m}\Theta^{2}+\Theta+\tau_{r}^{-1}I\right)
%          \Phi_{t}
%       \left(\tau_{m}\Theta^{2}+\Theta+\tau_{r}^{-1}I\right)
%          A_{t}^{-1} 
%       \\
    \bfeta &\equiv& 
      \left(\begin{array}{c}
         -1\\ - \tau_{m}\vartheta \\ 
         1\\ \tau_{m}\vartheta
      \end{array}\right),  
\label{MatriVarpi1}
\end{eqnarray}
respectively, 
with the $4\times 4$ matrix $\Phi_{t}$
defined by 
\begin{eqnarray}
   \Phi_{t} &\equiv& \int_{t_{0}}^{t}ds\; 
      \mathbf{K}_{s} \mathbf{K}_{s}^{T} .
      \label{FunctPhi1}
\end{eqnarray}
   Inserting Eq. (\ref{ActioFunct2}) into Eq. (\ref{EFunctWork3}) 
we obtain 
\begin{widetext}
\begin{eqnarray}
    \mathcal{E}_{w}(\lambda,t) &=& 
       C_{\mathcal{E}} 
       e^{ - \lambda(1-\lambda) \alpha\beta v^{2} (t-t_{0})}
       \int\!\int dx_{i}dp_{i} \int\!\int dx_{f}dp_{f} 
       f(x_{i},p_{i},t_{0}) 
       \nonumber \\
   &&\spaEq\times\exp\left\{
      -\frac{1}{4}\alpha\beta 
      \left[\mathbf{B}_{if}^{(1-2\lambda)}\right]^{T}
      \Lambda_{t} \mathbf{B}_{if}^{(1-2\lambda)}
      - \alpha\beta\lambda v \bfeta^{T}\mathbf{B}_{if}^{(0)} 
      \right\} .
      \nonumber\\
\label{EFunctWork4}
\end{eqnarray}
%
%   A derivation of Eq. (\ref{EFunctWork4}) is given in Appendix 
%\ref{WorkAppen2}. 
   Equation (\ref{EFunctWork4}) gives a concrete form of 
the function $\mathcal{E}_{w}(\lambda,t)$ 
for any initial distribution function $f(x_{i},p_{i},t_{0})$. 

   The $\lambda$-independent normalization coefficient  
$C_{\mathcal{E}}$ in Eq. (\ref{EFunctWork4}) can be determined 
from the condition (\ref{IdentE1}), and we obtain 
\begin{eqnarray}
   C_{\mathcal{E}} &=& \left\{
   \int\!\int dx_{i}dp_{i} \int\!\int dx_{f}dp_{f} 
       f(x_{i},p_{i},t_{0}) 
   %   \right.\nonumber \\
   %&&\spaEq\spaEq\left.\times
    \exp\left\{
      -\frac{1}{4}\alpha\beta \left[\mathbf{B}_{if}^{(1)}
      \right]^{T}
      \Lambda_{t} \mathbf{B}_{if}^{(1)} \right\}
   \right\}^{-1} .
\label{NormaConst1}
\end{eqnarray}
   Note that by using the condition (\ref{IdentE1}) 
we avoided to carry out explicitly the path integral  
included originally in the quantity $C_{\mathcal{E}}$  
[cf. Eq. (\ref{EFunctWork3})]. 

    Inserting Eq. (\ref{EFunctWork4}) into Eq. (\ref{DistriWork2}), 
and carrying out the Gaussian integral over $\lambda$ 
appearing then in Eq. (\ref{DistriWork2}),  
we obtain  
\begin{eqnarray}
   P_{w}(W,t) &=& \frac{ C_{\mathcal{E}}}{\sqrt{4
      \pi\alpha\beta v^{2} \left( t-t_{0} 
      -\tau_{r}^{2}\mathbf{J}^{T}\Lambda_{t}\mathbf{J}\right) }}
      \int\!\int dx_{i}dp_{i} \int\!\int dx_{f}dp_{f} \;
      f(x_{i},p_{i},t_{0}) 
      \nonumber \\
   && \spaEq\times
      \exp\left\{ -\frac{1}{4}\alpha\beta 
      \left[\mathbf{B}_{if}^{(1)}\right]^{T}
      \Lambda_{t} \mathbf{B}_{if}^{(1)}
  %    \right\}
  %    \nonumber \\
  % &&  \times
  %    \exp\left\{ 
      - \frac{\left\{W -\alpha\beta v\left[v(t-t_{0}) 
      + \left(\bfeta^{T}-\tau_{r}\mathbf{J}^{T}\Lambda_{t}\right)
      \mathbf{B}_{if}^{(1)}\right] 
      \right\}^{2}}
      {4\alpha\beta v^{2} \left( t-t_{0} 
      -\tau_{r}^{2}\mathbf{J}^{T}\Lambda_{t}\mathbf{J}\right)}
      \right\} 
   %  \nonumber \\
\label{DistriWork6}
\end{eqnarray}
\end{widetext}
where the 4-dimensional vector $\mathbf{J}$ is defined by 
\begin{eqnarray}
   \mathbf{J} \equiv 
      \left(\begin{array}{c}
      1 \\ 0  \\ 1 \\ 0  
   \end{array}\right) 
\label{VectorJ1}
\end{eqnarray}
and we used the relation $\bfeta^{T}\mathbf{J} = 0$. 
   Equation (\ref{DistriWork6}) is an explicit form for the work distribution function for all time, 
and for any initial distribution function 
$f(x_{i},p_{i},t_{0})$. %at the initial time $s=t_{0}$. 
   Using Eq. (\ref{NormaConst1}) for the coefficient 
$C_{\mathcal{E}}$, the work distribution function 
(\ref{DistriWork6}) is properly normalized, namely 
$\int dW\; P_{w}(W,t) =1$, at any time $t$.  

   In the next two sections \ref{AsympFluct} and \ref{InertEffec} 
we discuss, using the work distribution function  (\ref{DistriWork6}), 
fluctuation properties of the work  
from the viewpoint of the asymptotic fluctuation theorem 
for $t\rightarrow +\infty$, as well as for finite times.

%%%%%%%%%%%%%%%%%%%%%%%%%%%%%%%%%%%%%%%%%%%%%%%%%%%%%%%%%%%%%%%%%%%%%%
\section{Asymptotic Fluctuation Theorem}
\label{AsympFluct}

   The matrix $\Lambda_{t}$ defined by 
Eq. (\ref{MatriLambd1}) satisfies the condition 
\begin{eqnarray}
   \lim_{t\rightarrow +\infty}\frac{1}{t-t_{0}} \Lambda_{t} = 0 , 
\label{AsympCondi1}
\end{eqnarray}
as shown in Appendix \ref{AsympWork}. 
   Equation (\ref{AsympCondi1}) implies that 
$v(t-t_{0})+(\bfeta^{T}-\tau_{r}\mathbf{J}^{T}\Lambda_{t})
\mathbf{B}_{if}^{(1)} $ $
\stackrel{t\rightarrow +\infty}{\sim} v(t-t_{0})$ and 
$t-t_{0} -\tau_{r}^{2}\mathbf{J}^{T}\Lambda_{t}\mathbf{J}  
\stackrel{t\rightarrow +\infty}{\sim} t-t_{0}$ 
in Eq. (\ref{DistriWork6}), 
so that the work distribution function 
$P_{w}(W,t)$ is proportional to 
the Gaussian function 
$\exp\{-[W -\alpha\beta v^{2}(t-t_{0})]^{2}
/[4\alpha\beta v^{2}  (t-t_{0}) ]\}$ in the long time limit 
$t\rightarrow +\infty$, i.e.      
\begin{eqnarray}
   P_{w}(W,t) &\stackrel{t\rightarrow +\infty}{\sim}&
      \frac{1}{\sqrt{4
      \pi\alpha\beta v^{2} \left( t-t_{0}\right) }} 
         \nonumber \\
   &&\times 
   \exp\left\{-\frac{\left[W -\alpha\beta v^{2}\left(t-t_{0} 
      \right)\right]^{2}}
      {4\alpha\beta v^{2}  (t-t_{0})} \right\} 
      \spaEq
\label{DistriWork5}
\end{eqnarray}
regardless of the initial distribution function 
$f(x_{i},p_{i},t_{0})$. 
   It is important to note that the work distribution function 
(\ref{DistriWork5}) in the long time limit $t\rightarrow+\infty$
in the inertial case is the same as in the over-damped case. 
   Physically, this is, of course, due to the finiteness 
of the inertial characteristic time $\tau_{m}$, which makes 
inertial effects disappear in the long time limit. 
%   Physically, the over-damped assumption is justified  
%to discuss a long time behavior of the system, 
%so that results under the over-damped assumption 
%should be equivalent to those with the  
%inertia case in the long time limit. 
   Nevertheless, the proof of this equivalence is non-trivial.  
    
   From Eq. (\ref{DistriWork5}) we immediately derive   
\begin{eqnarray}
   \lim_{t\rightarrow +\infty}
   \frac{P_{w}(W,t)}{P_{w}(-W,t)} = e^{W}
\label{AsympFT1}
\end{eqnarray}
\emph{for any initial distribution function} $f(x_{i},p_{i},t_{0})$. 
   We will call Eq. (\ref{AsympFT1}) the asymptotic fluctuation 
theorem for work. 
   Equation (\ref{AsympFT1}) is independent of the value 
of the parameter $\vartheta$, i.e. of the frame of reference 
(laboratory or comoving) or also of the contribution of 
the d'Alembert-like force to the work (\ref{Work1}).

%%%%%%%%%%%%%%%%%%%%%%%%%%%%%%%%%%%%%%%%%%%%%%%%%%%%%%%%%%%%%%%%%%%%%%
\section{Inertial Effects for Finite Times}
\label{InertEffec}

\subsection{Slope of $\ln [P_{w}(W,t)/P_{w}(-W,t)]$ and the Critical Mass}
   
   In contrast to the asymptotic work distribution function 
(\ref{DistriWork5}), various inertial effects in 
the work distribution function appear for finite times. 
   In this section we discuss such inertial effects 
using the function $G(W,t)$ defined by 
\begin{eqnarray}
   G(W,t) \equiv \frac{\partial}{\partial W}
   \ln \frac{P_{w}(W,t)}{P_{w}(-W,t)} . 
\label{FunctG1}
\end{eqnarray}
   The function $G(W,t)$ gives the slope of the fluctuation 
function $\ln [P_{w}(W,t)$ $/P_{w}(-W,t)]$ with respect to $W$, 
and satisfies  
\begin{eqnarray}
   \lim_{t\rightarrow+\infty}G(W,t) = 1 
\label{AsympFT2}
\end{eqnarray}
by the asymptotic fluctuation theorem (\ref{AsympFT1}).%
\footnote{
   A function like $\tilde{G}(W,t) \equiv (1/\langle W\rangle ) 
\ln [P_{w}(W,t)/P_{w}(-W,t)]$ with the average work 
$\langle W\rangle$ has been used to characterize   
fluctuation theorems \cite{ZC03a}.  
   The function (\ref{FunctG1}) is connected 
to $\tilde{G}(W,t)$ by  
$G(W,t) = \langle W\rangle \partial \tilde{G}(W,t) /\partial W$. 
   One of the advantage to use $G(W,t)$ 
instead of $\tilde{G}(W,t)$ is that 
different from $\tilde{G}(W,t)$, $G(W,t)$ is independent 
of $W$ when the distribution function $P_{w}(W,t)$ is Gaussian, 
as shown in Eq. (\ref{DistriWork9}).   
}
 
   The behavior of $G(W,t)$ for finite times  
depends on the initial condition. 
   To get concrete results, in this section 
we concentrate on the case of a nonequilibrium steady state 
initial condition, which can be represented by 
\begin{widetext}
\begin{eqnarray}
   f(x_{i},p_{i},t_{0}) 
      = \frac{\beta}{2\pi} \sqrt{\frac{\kappa}{m}}
      \exp\left\{-\beta\left[
      \frac{(p_{i}-mv)^{2}}{2m}
      +\frac{1}{2}\kappa (x_{i}-vt_{0}+v\tau_{r})^{2}\right]\right\} 
      \label{InitiNNSS1} 
\end{eqnarray}
for any frame.  
   The initial distribution function (\ref{InitiNNSS1}) 
gives a Gaussian distribution for the particle 
initial position $x_{i}$ and momentum $p_{i}$ 
around their nonequilibrium steady state average values 
$vt_{0}-v\tau_{r}$ and $mv$, respectively. 
   Inserting Eq. (\ref{InitiNNSS1}) into Eq. (\ref{DistriWork6})  
the work distribution function is given by  
\begin{eqnarray}
   P_{w}(W,t) 
   &=& \sqrt{\frac{1-\Omega_{t}}
      {4\pi\alpha\beta v^{2} \left( t-t_{0} 
      -\tau_{r}^{2}\mathbf{J}^{T}\Lambda_{t}\mathbf{J}\right)}}
   %   \nonumber\\
   %&&\times
      \exp\left\{ -\frac{
      1- \Omega_{t}  }
      {4\alpha\beta v^{2} \left( t-t_{0} 
      -\tau_{r}^{2}\mathbf{J}^{T}\Lambda_{t}\mathbf{J}\right)}
      \left[ W -\alpha\beta v^{2}(t-t_{0}) \right]^{2}\right\} 
   %   \nonumber\\
\label{DistriWork9}
\end{eqnarray}
where $\Omega_{t}$ is defined by 
\begin{eqnarray}
   \Omega_{t} &\equiv& 
   \left(\bfeta-\tau_{r}\Lambda_{t}\mathbf{J}\right)^{T}
      \left[
      \left(\bfeta-\tau_{r}\Lambda_{t}\mathbf{J}\right)
          \left(\bfeta-\tau_{r}\Lambda_{t}\mathbf{J}\right)^{T}
   %    \right.\nonumber \\
   % &&\spaEq\left.
          +\left( t-t_{0} 
      -\tau_{r}^{2}\mathbf{J}^{T}\Lambda_{t}\mathbf{J}\right) 
      \left( \Lambda^{(0)}+\Lambda_{t} \right) 
      \right]^{-1}
      \left(\bfeta-\tau_{r}\Lambda_{t}\mathbf{J}\right) .
\label{FunctOmega1}
\end{eqnarray}
\end{widetext}
with the $4\times 4$ matrix $ \Lambda^{(0)}$ defined by 
\begin{eqnarray}
   \Lambda^{(0)} \equiv 
      \frac{2}{\alpha} 
      \left(\begin{array}{cccc}
      \kappa & 0 & 0 & 0 \\
      0      & m & 0 & 0 \\
      0      & 0 & 0 & 0 \\
      0      & 0 & 0 & 0  
   \end{array}\right) .
\label{TildeLambda1}
\end{eqnarray}
   [See Appendix \ref{FinitWork} for a derivation of Eq. 
(\ref{DistriWork9}).]
%   One may also notice a similarity between the matrix 
%$\Lambda^{(0)}$ and the asymptotic form (\ref{MatriLambd4}) of 
%the matrix $\Lambda_{t}$.] 
   Note that the work distribution function (\ref{DistriWork9}) 
is Gaussian with the average work $\langle W \rangle = 
\alpha\beta v^{2}(t-t_{0})$ at any time because we chose 
a Gaussian nonequilibrium steady state initial condition 
(\ref{InitiNNSS1}).   
   Since the work distribution function $P(W,t)$ is Gaussian, 
$G(W,t)$ defined by Eq. (\ref{FunctG1}) is independent 
of $W$, so that we denote it by $G(t) [=G(W,t)]$ from now on. 
   Inserting Eq. (\ref{DistriWork9}) into Eq. (\ref{FunctG1}), 
we obtain  
\begin{eqnarray}
   G(t) = \frac{1-\Omega_{t}}
   {1 -\frac{\tau_{r}^{2}}{t-t_{0}}\mathbf{J}^{T}
   \Lambda_{t}\mathbf{J}}
\label{FunctG2}
\end{eqnarray}
as an explicit form of $G(t)$. 
   One may notice that $G(t)$ in Eq. (\ref{FunctG2}) is 
independent of the dragging velocity $v$ and 
the inverse temperature $\beta$. 
   Moreover, $G(t)$ is positive for $t>t_{0}$ 
because the distribution function 
$P_{w}(W,t)$ is normalizable so that the coefficient  
$(1- \Omega_{t})/[4\alpha\beta v^{2} ( t-t_{0} 
-\tau_{r}^{2}\mathbf{J}^{T}\Lambda_{t}\mathbf{J})] 
=G(t)/[4\alpha\beta v^{2}( t-t_{0})]$ in the exponent 
of the Gaussian distribution function (\ref{DistriWork9}) 
must be positive.  

   As a first approximation to the asymptotic relaxation of $G(t)$ 
to its final value (\ref{AsympFT2}), 
we obtain from Eq. (\ref{FunctG2})   
\begin{eqnarray}
   G(t) \stackrel{t\rightarrow+\infty}{\sim}
      1+\frac{\tau_{r}-\tau_{m}\vartheta^{2}}
      {t-t_{0}-\tau_{r} + \tau_{m}\vartheta^{2}} ,
\label{FunctGAsym1}
\end{eqnarray}
meaning that the function $G(t)$ decays 
to $1$ by a power inversely proportional to the time   
in the long time limit $t\rightarrow +\infty$. 
[See Appendix \ref{AsympFunctG} for a derivation of Eq. (\ref{FunctGAsym1}).] 
   Equation (\ref{FunctGAsym1}) is only 
the first approximation for an asymptotic form 
of $G(t)$, but already includes an important inertial contribution  
to $G(t)$, as well as an interesting frame dependence of $G(t)$. 
   Actually, the second term on the right-hand side of Eq. 
(\ref{FunctGAsym1}) depends on the mass $m$ 
via $\tau_{m}=m/\alpha$ in the laboratory frame 
$\vartheta=1$, while that term is independent 
of the mass in the comoving frame $\vartheta=0$. 
   Another interesting property of $G(t)$ expressed 
by Eq. (\ref{FunctGAsym1}) is that  
in the laboratory frame $\vartheta=1$ the second term on  
the right-hand side of Eq. (\ref{FunctGAsym1}), 
the $t^{-1}$-decay term of $G(t)$,   
vanishes in the case that $\tau_{r} = \tau_{m}$, i.e. 
for a special mass value $m=\alpha^{2}/\kappa$.

%---------------------------------------------------------------------
\begin{figure}[!t]
\vspfigA
\begin{center}
\includegraphics[width=\widthfigB]{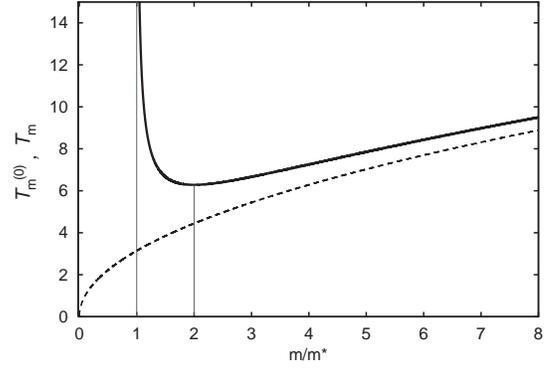}
\vspfigB
\caption{
      Time oscillation period $\mathcal{T}_{m}$ 
   (solid line) in $\tilde{x}_{s}^{*}$  
   as a function of mass $m$ normalized by the critical mass 
   $m^{*}\equiv \alpha^{2}/(4\kappa)$ 
   for the mass in $m/m^{*}>1$.  
      There is no time-oscillation for $m/m^{*}<1$ 
   and the minimum of $\mathcal{T}_{m}$ is at $m/m^{*}=2$. 
      Here, we used parameter values $\alpha =\kappa = 1$.  
      We also plotted a time-oscillation period 
   $\mathcal{T}_{m}^{(0)} =\left.\mathcal{T}_{m}\right|_{\alpha=0} 
   (= 2\pi \sqrt{m/\kappa})$
   (broken line) for a purely harmonic oscillation  
   in the case without the dissipation ($\alpha = 0$). 
      The time-oscillation period $\mathcal{T}_{m}$ 
   approaches $\mathcal{T}_{m}^{(0)}$ in the large mass limit 
   $m/m^{*}\rightarrow +\infty$. 
      }
\label{fig1oscilTime}
\end{center}
\vspfigC
\end{figure}  
%---------------------------------------------------------------------

   Perhaps the most interesting implication of Eq. (\ref{FunctG2}) 
for $G(t)$, although it does not appear explicitly in  
the asymptotic expression (\ref{FunctGAsym1}) of $G(t)$, 
is the existence of a critical value of the mass 
$m=m^{*}$ above which $G(t)$ shows a time-oscillatory behavior. 
   In our theory, this time-oscillation has its origin  
in the time-dependence of $\tilde{x}_{s}^{*}$
given by Eq. (\ref{SolutTildeX1}) via the exponential 
terms $\exp(\nus{\pm} t)$, etc., when the coefficient 
$\nus{\pm}$ given by Eq. (\ref{CoeffNuPM}) has 
an imaginary part, namely when the condition 
\begin{eqnarray}
   m > m^{*}\equiv \frac{\alpha^{2}}{4\kappa}
\label{CritiMass1}
\end{eqnarray}
(derived from the condition $4 \tau_{m}\tau_{r}^{-1}-1>0$) 
is satisfied. 
   We call the mass $m^{*}$ the critical mass in this paper,  
since a (smooth) ``dynamical'' phase transition  
takes place at $m=m^{*}$. 
   For masses $m>m^{*}$,   
the position $\tilde{x}_{s}^{*}$ has a time-oscillation 
with the oscillation period $\mathcal{T}_{m}$ 
\begin{eqnarray}
   \mathcal{T}_{m} = 
%   \frac{2\pi}{\sqrt{\kappa}}
%   \frac{m}{\sqrt{m-m^{*}}} 
%   = 2\pi \sqrt{\frac{m}{\kappa}
%   \frac{m}{m-m^{*}}} =
    2\pi \sqrt{\frac{m}{\kappa}}
   \left(1-\frac{m^{*}}{m}\right)^{-1/2}
\label{OscilPerio1}
\end{eqnarray}
corresponding to a frequency $\omega\equiv
\sqrt{4\tau_{m}\tau_{r}^{-1}-1}/(2\tau_{m})=|Im\{\nus{\pm}\}|$, 
using the imaginary part $Im\{\nus{\pm}\}$ of $\nus{\pm}$. 
   In Fig. \ref{fig1oscilTime} the time-oscillation 
period $\mathcal{T}_{m}$ (solid line) 
is shown as a function of the scaled mass 
$m/m^{*}$ for $m/m^{*}>1$.  
   There is no time-oscillation of $\tilde{x}_{s}^{*}$ 
in the case of $m/m^{*}<1$, and the time-oscillation period 
diverges when $m/m^{*}\rightarrow 1+0$.  
   The oscillation period $\mathcal{T}_{m}$ 
decreases rapidly as a function of mass $m$ 
for $m/m^{*} < 2$, has a minimum at $m/m^{*}=2$, 
and increases gradually for $m/m^{*}>2$. 
   For comparison, we also plotted in Fig. \ref{fig1oscilTime} 
the scaled mass dependence 
of the time-oscillation period $\mathcal{T}_{m}^{(0)}
= 2\pi \sqrt{m/\kappa} \; 
(=\left.\mathcal{T}_{m}\right|_{\alpha=0})$ (broken line) 
for a purely harmonic oscillator with   
spring constant $\kappa$. 
   Different from the time-oscillation period (\ref{OscilPerio1}),  
the period $\mathcal{T}_{m}^{(0)}$ is defined for all the masses, 
and increases monotonically as the mass increases.  
   The time-oscillation period (\ref{OscilPerio1})  
approaches $\mathcal{T}_{m}^{(0)}$ in the large mass limit 
$m/m^{*}\rightarrow +\infty$. 

   It is useful to consider the critical behavior 
in the time-oscillating behavior of $G(t)$  
as due to the presence of \emph{two} independent time scales 
appearing in our model:  
one characterized by $\tau_{r} (=\alpha/\kappa)$ 
and another by $\tau_{m} (=m/\alpha)$.   
   These time scales $\tau_{m}$ and $\tau_{r}$ are related by  
$\tau_{r}=4\tau_{m^{*}}$ at the critical mass $m=m^{*}$. 
   Using these two time scales, the time oscillation period 
$\mathcal{T}_{m}^{(0)}$ for a purely harmonic oscillator 
is given by $\mathcal{T}_{m}^{(0)}=2\pi \sqrt{\tau_{r} \tau_{m}}$. 
   Introducing the frequencies 
$\omega^{(0)}\equiv 2\pi / \mathcal{T}_{m}^{(0)}$ and
$\omega_{m}\equiv 1 / \tau_{m}$ corresponding to 
the two time scales $\mathcal{T}_{m}^{(0)}$ and $\tau_{m}$, 
respectively, 
the frequency $\omega \equiv 2\pi /\mathcal{T}_{m}$ 
is represented as 
$\omega = \sqrt{[\omega^{(0)}]^{2}-\omega_{m}^{2}/4}$ 
corresponding to the time-oscillation period (\ref{OscilPerio1}). 
   In this expression for the frequency $\omega$ 
the time oscillations occur only when 
the condition $[\omega^{(0)}]^{2}>\omega_{m}^{2}/4$ is satisfied. 
   The existence of these two time scale $\tau_{m}$ and 
$\tau_{r}$ is therefore essential for the time-oscillatory 
behavior with the frequency $\omega$, noting that there is no
time-oscillation in the over-damped case 
containing only $\tau_{r}$. %but not $\tau_{m}$. 
  
   In the next two subsections \ref{Vartheta0} and \ref{Vartheta1}, 
we investigate properties of $G(t)$ in more detail, 
including its time-oscillating behavior, for  
(A) the work done in the laboratory frame ($\vartheta = 1$), and 
(B) the work done in the comoving frame ($\vartheta = 0$), separately. 
   We will also compare those results with those for the 
over-damped case.  
   For this purpose, we now calculate $G(t)$ explicitly 
in the over-damped case. 
   In our previous paper \cite{TC07a}, we already calculated 
the work distribution function $P_{w}^{(0)}(W,t)$ 
for the over-damped case, which is given by 
\begin{eqnarray}
   P_{w}^{(0)}(W,t) &=& \frac{1}{\sqrt{4\pi\alpha \beta v^{2} 
      \left[ t-t_{0}-\tau_{r}(1-b_{t}) \right] }}
   \nonumber \\
   &&\spaEq\times
      \exp\left\{-\frac{\left[W-\alpha\beta v^{2}
      (t-t_{0})\right]^{2}}{4\alpha\beta v^{2}\left[ t-t_{0}
      -\tau_{r}(1-b_{t}) \right]}\right\} 
      \nonumber \\
\label{WorkDistr4}
\end{eqnarray}
with $b_{t}\equiv \exp[-(t-t_{0})/\tau_{r}]$  
in the case of a nonequilibrium steady state initial distribution 
function $f^{(0)}(x_{i},t_{0}) = \sqrt{\beta\kappa/(2\pi)}$ $
\exp[-\beta \kappa (x_{i}-vt_{0}+v\tau_{r})^{2}/2]$ 
for the particle position $x_{i}$ for the over-damped case  
at the initial time $t_{0}$.%
\footnote{ 
   Note that the work distribution function (\ref{WorkDistr4}) 
approaches the distribution function (\ref{DistriWork5}) 
in the long time limit $t\rightarrow +\infty$ 
because of $t-t_{0}-\tau_{r}(1-b_{t})  
\stackrel{t\rightarrow +\infty}{\sim} t-t_{0}$.
} 
   Using Eq. (\ref{WorkDistr4}), and defining, 
[cf. Eq. (\ref{FunctG1})], 
$G^{(0)}(t) \equiv (\partial/\partial W) 
\ln [P_{w}^{(0)}(W,t)/P_{w}^{(0)}(-W,t)]$,  
we have  
\begin{eqnarray}
   G^{(0)}(t) 
   = 1+ \frac{\tau_{r}(1-b_{t})}{t-t_{0}-\tau_{r}(1-b_{t})} ,
%  = \left[1-\frac{\tau_{r}(1-b_{t})}{t-t_{0}} \right]^{-1} . 
\label{FunctOD1}
\end{eqnarray}
which gives $G(t)$ for the over-damped case \cite{ZC03b}. 
   Note that Eq. (\ref{FunctOD1}) implies 
$G^{(0)}(t) \stackrel{t\rightarrow +\infty}{\sim}
1  + \tau_{r} /(t-t_{0}-\tau_{r})$, which is consistent with  
Eq. (\ref{FunctGAsym1}), since $\tau_{m}$ is zero  
for the over-damped case.

%%%%%%%%%%%%%%%%%%%%%%%%%%%%%%%%%%%%%%%%%%%%%%%%%%%%%%%%%%%%%%%%%%%%%%
\subsection{$G(t)$ in the Laboratory Frame}
\label{Vartheta0} 

   In this subsection we consider $G(t)$ 
given by Eq. (\ref{FunctG2}) 
$\;-$ 
which depends on the parameter $\vartheta$ to specify a frame 
via $\Lambda_{t}$ and $\Omega_{t}$ 
$-\;$ 
for the work done in the laboratory frame, 
i.e. for $\vartheta=1$. 
   In this subsection \ref{Vartheta0}, as well as in  
the next subsection \ref{Vartheta1}, we use the parameter 
values $\alpha =\kappa = 1$ and set the initial time $t_{0}=0$, 
i.e. $\tau_{r}$=1 as a time unit 
and $m/m^{*} %=4\tau_{m}/\alpha 
= 4\tau_{m}$ as the scaled mass. 

\begin{figure}[!t]
\vspfigA
\begin{center}
\includegraphics[width=\widthfigA]{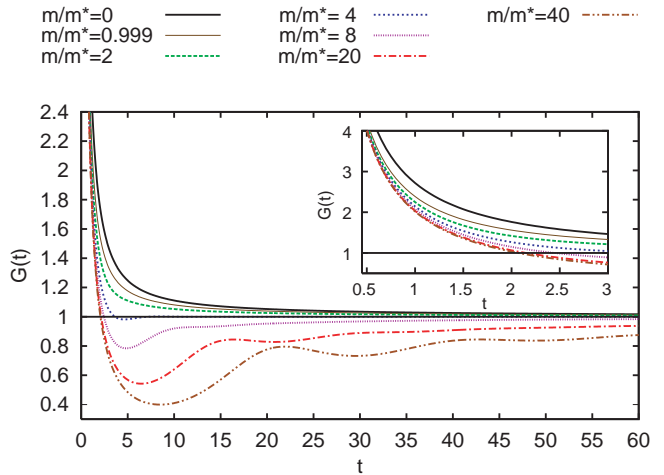}
\vspfigB
\caption{(color online)
   Graphs of $G(t)=(\partial/\partial W) 
   \ln [P_{w}(W,t)/P_{w}(-W,t)]$ as a function of time $t$ 
   for the work done in the laboratory frame ($\vartheta=1$) 
   in the case of a nonequilibrium steady state initial condition 
   for $t\in [0,60]$. 
      Inset: Graphs of $G(t)$ in a short time period for $t\in [0,3]$. 
      Lines in these graphs correspond to parameter values 
   of the scaled masses $m/m^{*}=0$ (over-damped case), 
   $0.999$, $2$, $4$, $8$, 
   $20$ and $40$ as indicated above this figure,  
   and we used parameter values $\alpha = \kappa =1$ 
   (so that $\tau_{r}$=1 for a unit time and also 
   $m/m^{*} = 4\tau_{m}$ as the scaled mass,  
   with $m^{*}=1/4$)  
   and $t_{0}=0$. 
      }
\label{fig2slopeG0}
\end{center}
\vspfigC
\end{figure}

   Figure \ref{fig2slopeG0} shows $G(t)$ 
given by Eq. (\ref{FunctG2}) 
as a function of time $t$ for the scaled masses $m/m^{*}=0$ 
(over-damped case), $0.999$, $2$, $4$, $8$, $20$ and $40$.   
   The graphs of $G(t)$ all converge to 1 in 
the long time limit $t\rightarrow+\infty$, 
as required by the asymptotic fluctuation theorem (\ref{AsympFT1}),   
i.e. by Eq. (\ref{AsympFT2}).

%---------------------------------------------------------------------
%---------------------------------------------------------------------

   We now discuss in some detail 
the properties of Fig. \ref{fig2slopeG0}. 
   This figure shows that $G(t)$ 
for nonzero masses is always smaller than 
in the over-damped case of zero mass. 
   In the over-damped case, $G(t)$ decreases monotonically 
to the final value $1$ from $+\infty$ at the initial time. 
   A similar behavior is still observed for small masses   
(e.g. see the graph for $m/m^{*}= 0.999$ in Fig. \ref{fig2slopeG0}.    
   It may also be noted that for small nonzero masses 
the relaxation of $G(t)$ to its final value $1$  
is faster than in the over-damped case  
(e.g. see the graphs for $m/m^{*}=2$ and $4$ 
in Fig. \ref{fig2slopeG0}). 
   This feature can be explained by the second term 
on the right-hand side of Eq. (\ref{FunctGAsym1}),  
since the absolute value 
$|\tau_{r}-\tau_{m}|$ of the numerator   
of this term is smaller  for $\vartheta = 1$ than 
the corresponding over-damped value $\tau_{r}$ 
in the case of $0<m/m^{*}< 8$, using that 
$|\tau_{r}-\tau_{m}|<\tau_{r}$. 
   Moreover, Fig. \ref{fig2slopeG0} shows that for large masses 
(e.g. see the graphs for $m/m^{*}>4$ in Fig. \ref{fig2slopeG0}),  
$G(t)$ is smaller than $1$ for long times, 
while $G(t)$ is always larger than $1$ in the over-damped case. 
   This is because the second term 
on the right-hand side of Eq. (\ref{FunctGAsym1}) 
is negative for $\tau_{r}<\tau_{m}$ (i.e. $m/m^{*}> 4$),    
when $\vartheta=1$ and $t>t_{0}+\tau_{r}-\tau_{m}$. 

%---------------------------------------------------------------------
\begin{figure}[!t]
\vspfigA
\begin{center}
\includegraphics[width=\widthfigA]{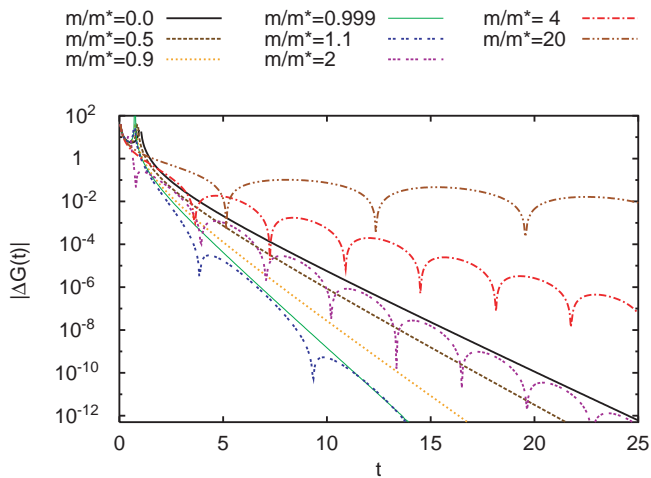}
\vspfigB
\caption{(color online)
      Linear-log plots of absolute value $|\Delta G(t)|$ 
    of the function $\Delta G(t) =  
   G(t)-1-(\tau_{r}-\tau_{m})/(t-\tau_{r}+\tau_{m})$ 
   as a function of time $t$ 
   for the work done in the laboratory frame  
   in the case of a nonequilibrium steady state initial condition. 
      Lines in these graphs correspond to parameter values 
   of the scaled masses $m/m^{*}=0$,  
   $0.5$, $0.9$, $0.999$,$1.1$, $2$, $4$ and $20$   
   as indicated above this figure.   
      The minima of the oscillations of  $|\Delta G(t)|$ 
   for $m/m^{*}>1$
   are actually zero, which is not indicated in 
   this figure and Figs. \ref{fig4oscilG0B}, 
   \ref{fig6oscilG1A} and \ref{fig7oscilG1B}. 
      We use the same parameter values 
   $\alpha$, $\kappa$ and $t_{0}$ as in Fig. \ref{fig2slopeG0}.   
      }
\label{fig3oscilG0A}
\end{center}
\vspfigC
\end{figure}  
%---------------------------------------------------------------------
%---------------------------------------------------------------------
\begin{figure}[!t]
\vspfigA
\begin{center}
\includegraphics[width=\widthfigA]{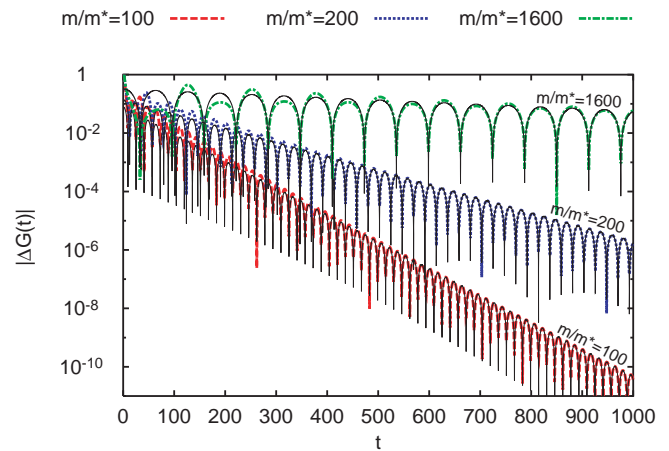}
\vspfigB
\caption{(color online)
      Long time behavior of $|\Delta G(t)|$ 
   as a function of time $t$ as linear-log plots 
   for the work done in the laboratory frame  
   in the case of a nonequilibrium steady state initial condition. 
      Here, we use the same parameter values 
   $\alpha$, $\kappa$ and $t_{0}$ as in Fig. \ref{fig2slopeG0}.   
      Broken, dotted and dash-dotted lines 
   in these graphs correspond 
   to parameter values 
   of the scaled masses $m/m^{*}=100$, $200$ and $1600$, 
   respectively. 
      Solid lines are fits of $|\Delta G(t)|$
   to the function (\ref{FitFunct1}) 
   using Table \ref{FittiParam1}, together 
   with the time-oscillation period (\ref{OscilPerio1}), 
   but they are visually indistinguishable 
   from the graphs of $|\Delta G(t)|$ 
   except for short times. 
      }
\label{fig4oscilG0B}
\end{center}
\vspfigC
\end{figure}  
%---------------------------------------------------------------------

   A time-oscillatory behavior 
of $G(t)$ is clearly visible in Fig. \ref{fig2slopeG0}  
for large masses, i.e. for $m >\!> m^{*}$.
   To show more clearly the time-oscillatory behavior of $G(t)$ 
for $m > m^{*}$ as opposed to for $m < m^{*}$,  
we plotted in Fig. \ref{fig3oscilG0A} 
the absolute value of the deviation%
\footnote{
   We note that in this subsection we use 
the function (\ref{DeltaG1}) 
for $\vartheta=1$, while in the next subsection \ref{Vartheta1}
we use the function (\ref{DeltaG1}) for $\vartheta=0$.
}  
\begin{eqnarray}
   \Delta G(t) \equiv G(t)-1
      -\frac{\tau_{r}-\tau_{m}\vartheta^{2}}
      {t-t_{0}-\tau_{r} + \tau_{m}\vartheta^{2}} 
\label{DeltaG1}
\end{eqnarray}
of $G(t)$ from its asymptotic form (\ref{FunctGAsym1}) 
as a function of time $t \in [0,25]$ for the cases of 
$m/m^{*}=0$, $0.5$, $0.9$, $0.999$, $1.1$, $2$, $4$ and $20$. 
   To illustrate the long time behavior of $|\Delta G(t)|$ 
in more detail, we also show in Fig. \ref{fig4oscilG0B} 
the absolute value $|\Delta G(t)|$ of $\Delta G(t)$ as 
functions of $t\in[0,1000]$ for the scaled masses 
$m/m^{*}=100$, $200$ and $1600$ as linear-log plots.  
   The deviation $\Delta G(t)$ 
goes to zero when $t\rightarrow +\infty$ 
because of the asymptotic fluctuation theorem 
(\ref{AsympFT2}). 
   In  Figs. \ref{fig3oscilG0A} and \ref{fig4oscilG0B}, 
it is important to note 
that there is no time-oscillation of $\Delta G(t)$ 
for $0\leq m/m^{*}<1$, while 
we do observe time-oscillations of $\Delta G(t)$ 
for $m/m^{*}>1$, in agreement with a critical mass 
(\ref{CritiMass1}), above which $G(t)$ oscillates in time. 
   The decay of $|\Delta G(t)|$ to zero as a function of $t$ 
is faster for larger masses for $0\leq m/m^{*}<1$ 
(cf. Fig. \ref{fig3oscilG0A}), but 
slower for larger masses  for $m/m^{*}>1$ 
(cf. Figs. \ref{fig3oscilG0A} and \ref{fig4oscilG0B}). 
   
   To check that the time oscillation 
period $\mathcal{T}_{m}$ given by Eq. (\ref{OscilPerio1}) 
indeed appears in $G(t)$,  
we fitted the data for $\Delta G(t)$ to the function 
\begin{eqnarray}
   \Delta G(t) \stackrel{t\rightarrow+\infty}{\sim} 
      a e^{-b t}\sin\left(\frac{2\pi}{\mathcal{T}_{m}}t+c\right)
\label{FitFunct1}
\end{eqnarray}
with fitting parameters $a$, $b$  and $c$ 
in Fig. \ref{fig4oscilG0B}. 
   The values of the fitting parameters $a$, $b$ and $c$ 
are given in Table \ref{FittiParam1}.  
   The function (\ref{FitFunct1}) is then sufficiently close to 
$\Delta G(t)$ over many time-oscillation periods  
(except for short times), 
%as in Fig. \ref{fig4oscilG0B}, 
to suggest that the time-oscillations 
of $G(t)$ may well have the same origin as those in 
the position $\tilde{x}_{s}^{*}$.  
   Similarly for Fig. \ref{fig3oscilG0A},  
using the fitting function (\ref{FitFunct1}) 
we can also check that the time-oscillation periods 
of $|\Delta G(t)|$ in this figure 
are given by Eq. (\ref{OscilPerio1}).  
   We fully realize that Figs. \ref{fig3oscilG0A} 
and \ref{fig4oscilG0B} 
are not enough to specify convincingly the function form 
of decay of $\Delta G(t)$.  
   In Eq. (\ref{FitFunct1}) we assumed an exponential decay 
by a factor $a \exp(-b t)$, which seems to fit reasonably well 
the data in Fig. \ref{fig4oscilG0B}.   
   However, values of the fitting parameters $a$ and $b$ 
shown in Table \ref{FittiParam1} appear to vary non-negligibly 
if we fit data including longer time periods than the ones shown 
in Fig. \ref{fig4oscilG0B}. 
   In this sense, at this stage, the exponential factor  
in Eq. (\ref{FitFunct1}) 
should be regarded only as a convenience to check numerically 
the time oscillation period $\mathcal{T}_{m}$ appearing 
in $\Delta G(t)$, rather than claiming an asymptotic exponential 
decay of $\Delta G(t)$ of the form (\ref{FitFunct1}).

%---------------------------------------------------------------------
\begin{table*}[!t]
\begin{center}
\begin{tabular}{c|c|c|c|c|c}
    \makebox[\widthtableB]{Frame ($\vartheta$)} & 
    \makebox[\widthtableA]{$m/m^{*}$} & 
    \makebox[\widthtableA]{$\mathcal{T}_{m}$} & 
    \makebox[\widthtableA]{a} & 
    \makebox[\widthtableA]{b} & 
    \makebox[\widthtableA]{c} \\
 \hline
 \hline
% Laboratory (1) &   50 &  22.4 & -0.17  & 0.044  & 4.3  \\
 Laboratory (1) &  100 &  31.6 & -0.077 & 0.021  & 4.4  \\
 Laboratory (1) &  200 &  44.5 & -0.14  & 0.011  & 4.5  \\
% Laboratory (1) &  400 &  62.9 & -0.22  & 0.0060 & 4.6 \\
% Laboratory (1) &  800 &  88.9 & -0.30  & 0.0033 & 4.6 \\
 Laboratory (1) & 1600 & 125.7 & -0.32  & 0.0017 & 4.6  \\
% Laboratory (1) & 3200 & 177.7 & -0.28  & 0.00080 & 4.7  \\
 \hline
% Comoving (0) &   50 &  22.4 & 0.020  & 0.044  & 4.6   \\
 Comoving (0) &  100 &  31.6 & 0.0038 & 0.021  & 4.6  \\
 Comoving (0) &  200 &  44.5 & 0.0035 & 0.011  & 4.6  \\
% Comoving (0) &  400 &  62.9 & 0.0038 & 0.0064 & 4.7  \\
% Comoving (0) &  800 &  88.9 & 0.0038 & 0.0039 & 4.7 \\
 Comoving (0) & 1600 & 125.7 & 0.0032 & 0.0024 & 4.7 \\
% Comoving (0) & 3200 & 177.7 & 0.00019 & 0.00069 & 4.7 \\
 \hline
\end{tabular}
\end{center}
\caption{
   Values of the fitting parameters $a$, $b$ and $c$ for 
   the function (\ref{FitFunct1}) plotted 
   in Figs. \ref{fig3oscilG0A} and \ref{fig6oscilG1A}  
   for the parameter values $\alpha = \kappa =1$.  
%   so that $\tau_{r}=1$.
} 
\label{FittiParam1}
\vspfigC
\end{table*}% 
%---------------------------------------------------------------------

%%%%%%%%%%%%%%%%%%%%%%%%%%%%%%%%%%%%%%%%%%%%%%%%%%%%%%%%%%%%%%%%%%%%%%
\subsection{$G(t)$ in the Comoving Frame}
\label{Vartheta1}

   Here we consider $G(t)$ for the work done in  
the comoving frame, namely the case of $\vartheta=0$, 
in which the work includes effects of an inertial or 
d'Alembert-like force. 

%---------------------------------------------------------------------
\begin{figure}[!t]
\vspfigA
\begin{center}
\includegraphics[width=\widthfigA]{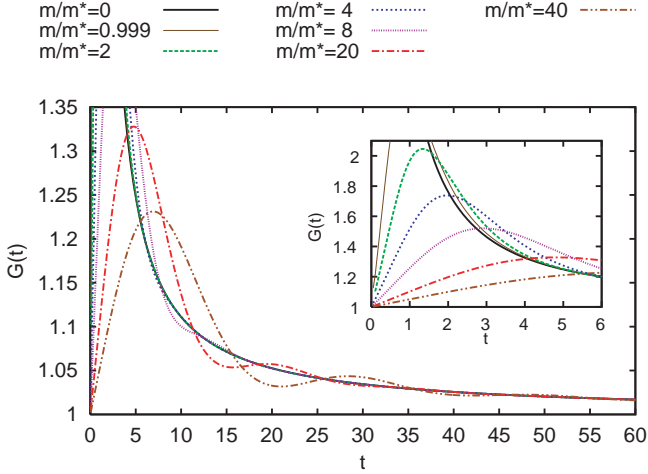}
\vspfigB
\caption{(color online)
      Graphs of $G(t)=(\partial/\partial W) 
   \ln [P_{w}(W,t)/P_{w}(-W,t)]$ as a function of time $t$ 
   for the work done in the comoving frame ($\vartheta=0$) 
   in the case of a nonequilibrium steady state initial condition 
   for $t\in [0,60]$. 
      Inset: Graphs of $G(t)$ in a short time period for $t\in [0,6]$. 
      Lines in these graphs correspond to parameter values 
   of the scaled masses $m/m^{*}=0$ (over-damped case),
   $0.999$, $2$, $4$, $8$, $20$ and $40$  
   as indicated above this figure  
   and we use the same parameter values 
   $\alpha$, $\kappa$ and $t_{0}$ as in Fig. \ref{fig2slopeG0}.   
      }
\label{fig5slopeG1}
\end{center}
\vspfigC
\end{figure}  
%---------------------------------------------------------------------

   Figure \ref{fig5slopeG1} shows graphs of $G(t)$ 
given by Eq. (\ref{FunctG2}) as a function of time $t$.  
   We chose the same masses as in Fig. \ref{fig2slopeG0}, 
namely $m/m^{*}=0$ (over-damped case), $0.999$, $2$, $4$, $8$, 
$20$ and $40$ with the critical mass $m^{*}=1/4$. 
   It is clear that in Fig. \ref{fig5slopeG1} 
graphs of $G(t)$ approach $1$ as $t\rightarrow+\infty$, 
confirming the asymptotic fluctuation theorem (\ref{AsympFT1}).
   
   Comparing Fig. \ref{fig2slopeG0} with Fig. \ref{fig5slopeG1},  
a dramatic difference in the behavior of $G(t)$ in the two 
frames is clearly visible. 
   First, a striking frame-dependence of $G(t)$ is that 
for any nonzero mass, $G(t)$ in the comoving frame starts 
from a finite value at the initial time $t_{0}(=0)$ 
and is always larger than $1$, 
in fact going through a maximum to its final value $1$. 
   This contrary to in the laboratory frame where 
$G(t)$ diverges for $t\rightarrow t_{0}+0$ and 
can be smaller than $1$ for large masses and long times 
as discussed in Sec. \ref{Vartheta0}. 
   Another remarkable point is that,  
different from in the laboratory frame 
as shown in Fig. \ref{fig2slopeG0},  
$G(t)$ converges to the over-damped line,  
much before converging to its final value $1$,   
as shown in Fig. \ref{fig5slopeG1}. 
   This feature can be explained by the asymptotic form 
(\ref{FunctGAsym1}) of $G(t)$, whose right-hand side is 
independent of the mass $m$ in the comoving frame ($\vartheta=0$),  
so a relaxation behavior of $G(t)$ to its final value $1$ 
in this frame should be close to that of the over-damped case.

%---------------------------------------------------------------------
\begin{figure}[!t]
\vspfigA
\begin{center}
\includegraphics[width=\widthfigA]{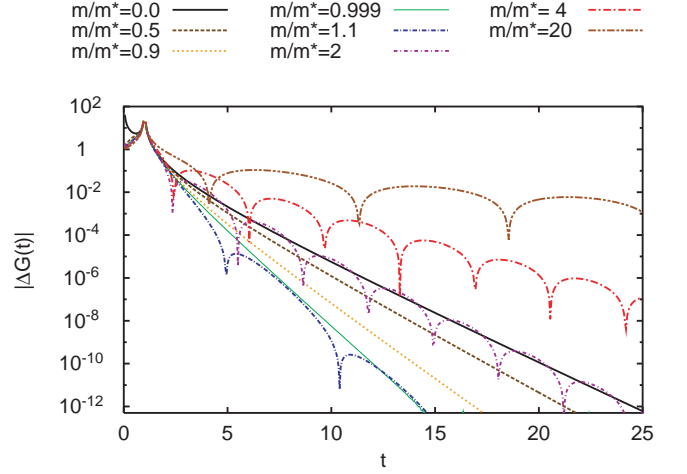}
\vspfigB
\caption{(color online)
      Linear-log plots of the absolute value $|\Delta G(t)|$ 
   of the function $\Delta G(t) = G(t)-1-\tau_{r}/(t-\tau_{r})$ 
   as a function of time $t$
   for the work done in the comoving frame  
   in the case of a nonequilibrium steady state initial condition. 
      Lines in these graphs correspond to parameter values 
   of the scaled masses $m/m^{*}=0$, 
   $0.5$, $0.9$, $0.999$,$1.1$, $2$, $4$ and $20$   
   as indicated above this figure   
   and we use the same parameter values 
   $\alpha$, $\kappa$ and $t_{0}$ as in Fig. \ref{fig2slopeG0}.   
      }
\label{fig6oscilG1A}
\end{center}
\vspfigC
\end{figure}  
%---------------------------------------------------------------------
%---------------------------------------------------------------------
\begin{figure}[!t]
\vspfigA
\begin{center}
\includegraphics[width=\widthfigA]{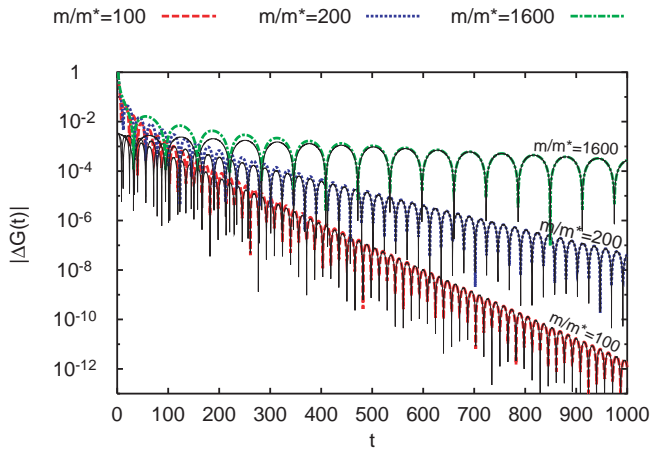}
\vspfigB
\caption{(color online)
      Long time behavior of $|\Delta G(t)|$ 
   as a function of time $t$ as linear-log plots 
   for the work done in the comoving frame  
   in the case of a nonequilibrium steady state initial condition. 
      Here, we use the same parameter values 
   $\alpha$, $\kappa$ and $t_{0}$ as in Fig. \ref{fig2slopeG0}.   
      Broken, dotted and dash-dotted lines 
   in these graphs correspond to parameter values 
   of the scaled masses $m/m^{*}=100$, $200$ and $1600$, 
   respectively. 
      Solid lines are fits of $|\Delta G(t)|$
   to the function (\ref{FitFunct1}) 
   using Table \ref{FittiParam1}, 
   and they are visually indistinguishable 
   from the graphs of $|\Delta G(t)|$ 
   except for short times.    
      }
\label{fig7oscilG1B}
\end{center}
\vspfigC
\end{figure}  
%---------------------------------------------------------------------

   Now, we discuss the  time-oscillatory behavior of $G(t)$ 
in the comoving frame. 
   We note that in the comoving frame the approach of $G(t)$ 
to its final value $1$ is via oscillations 
around the over-damped line, 
contrary to in the laboratory frame where 
this approach is unrelated to the over-damped line.  
   Such time-oscillations are already visible for large masses 
$m >\!> m^{*}$ in Fig. \ref{fig5slopeG1}, but 
to show them in a more magnified way, 
we plotted in Fig. \ref{fig6oscilG1A} the absolute value  
$|\Delta G(t)|$ of the function $\Delta G(t)$ defined 
by Eq. (\ref{DeltaG1}) as a function of time $t$ as linear-log plots.     
   Here, we plotted data for the scaled masses $m/m^{*}=0$, 
$0.5$, $0.9$, $0.999$, $1.1$, $2$, $4$ and $20$
and for the time period $t\in [0,25]$. 
   It is shown in Fig. \ref{fig6oscilG1A} that 
time-oscillations of $\Delta G(t)$ occur for $m/m^{*}>1$ 
but not for $0 \leq m/m^{*} <1$.  
   Moreover, $\Delta G(t)$ for $0 \leq m/m^{*} <1$ decays 
faster, while $\Delta G(t)$ for $m/m^{*}>1$ decays slower 
with time, for increasing mass. 
   These features are similar to those in the laboratory frame.

   In Fig. \ref{fig7oscilG1B} we show linear-log plots 
of $|\Delta G(t)|$ as functions of $t$ 
for longer times $t\in [0,1000]$ and for larger masses 
$m/m^{*}= 100, 200$ and $1600$ than in Fig. \ref{fig6oscilG1A}.  
   Comparing this figure in the comoving frame 
with the corresponding Fig. \ref{fig4oscilG0B} 
in the laboratory frame, we can see that 
the time-oscillation amplitudes of the function $\Delta G(t)$ 
in the comoving frame are much smaller than the corresponding 
ones in the laboratory frame, except for short times. 
   This should be noted as an important frame-dependence  
in the behavior of $G(t)$.  

   The time-oscillation periods appearing 
in Figs. \ref{fig6oscilG1A} and \ref{fig7oscilG1B} 
can be checked by fitting the data again  
to the function (\ref{FitFunct1}) 
with the time-oscillation period (\ref{OscilPerio1}). 
   We only show such fitting lines for Fig. \ref{fig7oscilG1B} 
using the fitting parameters $a$, $b$ and $c$ 
of Table \ref{FittiParam1}. 
   Like for the fitting lines in Fig. \ref{fig4oscilG0B}, 
the parameter values of $a$ and $b$ 
in Table \ref{FittiParam1} in the comoving frame 
also appear to vary non-negligibly for data for a longer time 
period than that shown in Fig. \ref{fig7oscilG1B}. 
   Therefore, as for Fig. \ref{fig4oscilG0B}, the fitting lines 
in Fig. \ref{fig7oscilG1B} should not be regarded as evidence 
for the exponential decay in the fitting function (\ref{FitFunct1}).  
   However, the fits of their time-oscillation periods 
of $|\Delta G(t)|$ to the function (\ref{FitFunct1}) 
in Fig. \ref{fig7oscilG1B} are satisfactory, 
which suggests again that the time-oscillations of $G(t)$ have 
the same origin as those in the position $\tilde{x}_{s}^{*}$, 
like in the laboratory frame.

%%%%%%%%%%%%%%%%%%%%%%%%%%%%%%%%%%%%%%%%%%%%%%%%%%%%%%%%%%%%%%%%%%%%%%
\section{Summary and Remarks} 
\label{ConclRemar}

   As a summary of this paper, we have discussed 
inertial effects related to the particle mass $m$ 
in nonequilibrium work distribution functions  
and their associated fluctuation theorems  
for a dragged Brownian particle model confined by a 
harmonic potential using 
a path integral approach for all times: asymptotic  
as well as finite. 
   We considered two kinds of work: 
the work $\mathcal{W}_{l}$ done in the laboratory frame 
and the work  $\mathcal{W}_{c}$ done in 
the comoving frame and we calculated the distribution 
functions $P_{w}(W,t)$ for them. 
   Using the distributions for the work in the different frames   
we analytically proved, for any initial condition, 
an asymptotic work fluctuation theorem, 
which has the same form in both the frames. 
   This contrasts with what happens for finite times, 
when for a nonequilibrium steady state initial condition   
there are major differences between the work fluctuations 
in the laboratory and comoving frames.    
   This was discussed, using  
the quantity $G(t) \equiv (\partial/\partial W) 
\ln [P_{w}(W,t)/P_{w}(-W,t)]$, which approaches 
the value $1$ in the long time limit $t\rightarrow +\infty$ 
by the asymptotic fluctuation theorem. 
   The $G(t)$ for the work $\mathcal{W}_{c}$ 
done in the comoving frame is larger than 1 at all times and 
converges to the corresponding over-damped value 
much before converging to its final value $1$.   
   On the other hand, the $G(t)$ for the work $\mathcal{W}_{l}$ 
done in the laboratory frame can be smaller than $1$ 
for sufficiently large times and masses, and the relaxation behavior 
of $G(t)$ to its final value $1$ is very different from that  
for the over-damped case, even for long times.  
   As one of the significant effects for finite times, 
we also discussed the existence of a critical mass $m^{*}$, 
so that for the mass $m > m^{*}$ 
a time-oscillatory behavior appears in $G(t)$ 
in both frames.

   In the remainder of this section, we make some remarks 
on the contents in the main text of this paper.

   1) We have discussed in this paper differences between 
the works $\mathcal{W}_{l}$ and $\mathcal{W}_{c}$,  
which originate in a frame dependence of 
the kinetic energy difference due to the d'Alembert-like force
as we discussed in Sec. \ref{WorkDragParti}.  
   In contrast to the work and the kinetic energy difference, 
the heat (as well as the potential energy difference) 
is frame-independent even in the inertial case.  
%but there is a difference in the kinetic energy difference 
%on the two frames, due to the difference in the  particle velocity. 
   Note that the two works $\mathcal{W}_{l}$ and $\mathcal{W}_{c}$ 
have the same average value in the nonequilibrium steady state,  
because their difference can be represented as 
a ``boundary term'' 
\begin{eqnarray}
   \mathcal{W}_{l} - \mathcal{W}_{c} 
   %= \int_{t_{0}}^{t} m\ddot{x}_{s} v 
   = m (\dot{x}_{t} -\dot{x}_{t_{0}} ) v 
%\label{}
\end{eqnarray}
depending on a difference between 
the two boundary values of $\dot{x}_{s}$ 
at the final time $s=t$ and the initial time $s=t_{0}$ only,  
so that the average of this boundary term 
$m (\dot{x}_{t} -\dot{x}_{t_{0}}) v$ is zero 
in the nonequilibrium steady state. 
   Nevertheless, this difference 
$m (\dot{x}_{t} -\dot{x}_{t_{0}} ) v$ between  
$\mathcal{W}_{l}$ and $\mathcal{W}_{c}$ causes  
dramatic differences in the work fluctuations, as shown in 
the subsections \ref{Vartheta0} and \ref{Vartheta1}
%Sec. \ref{InertEffec} 
of this paper.

%---------------------------------------------------------------------
\begin{table*}[!t]
\begin{center}
\begin{tabular}{c|c|c|c|c|c}
    \makebox[\widthtableD]{Brownian particle} & 
    \makebox[\widthtableC]{$x_{s}$} & 
    \makebox[\widthtableC]{$m$} & 
    \makebox[\widthtableC]{$\alpha$} & 
    \makebox[\widthtableC]{$\kappa$} & 
    \makebox[\widthtableC]{$\kappa v$} \\
 \hline 
    Torsion pendulum &
    $\theta_{s}$ & $I$ & $\nu$ & $C$ & $\mu$ \\
 \hline 
\end{tabular}
\end{center}
\caption{
      Correspondences  
   between the dragged Brownian particle model 
   described by Eq. (\ref{LangeEquat1}) and 
   the torsion pendulum model 
   described by Eq. (\ref{LangeEquat2}).
}
\label{CorreModel1}
\vspfigC
\end{table*}%
%--------------------------------------------------------------------- 

   2) In a different nonequilibrium model described 
by a linear Langevin equation, 
Ref. \cite{DJG06} considered the motion of a torsion pendulum 
under an external torque in a fluid. 
   The corresponding Langevin equation for the angular 
displacement $\theta_{s}$ of the pendulum at time $s$ 
in this system is then given by   
\begin{eqnarray}
   I \frac{d^{2} \theta_{s}}{ds^{2}} 
   = - \nu \frac{d \theta_{s}}{ds} 
   - C \theta_{s} + M_{s} + \zeta_{s}
\label{LangeEquat2}
\end{eqnarray}
where $I$ is the total moment of inertia of the displaced mass, 
$\nu$ is the viscous damping,  
$C$ the elastic torsional stiffness of the pendulum, 
$M_{s}$ the external torque, and $\zeta_{s}$ 
the Gaussian-white random force. 
   For this model, Ref. \cite{DJG06} considered 
the case of a linear torque of   
\begin{eqnarray} 
   M_{s} = \mu s
\label{LinearForci1}
\end{eqnarray}
with a force constant $\mu$.
   It is important to note that Eq. (\ref{LangeEquat2}) 
with the force (\ref{LinearForci1})
has mathematically the same form as the Langevin equation 
(\ref{LangeEquat1}) with the correspondences shown in 
Table \ref{CorreModel1}.  
   Based on these correspondences between the two models, 
for example, there should be a critical value $I^{*}$ of the total 
moment of inertia above which a similar time-oscillatory 
behavior occurs  in the pendulum model, 
like above the critical mass $m^{*}$ 
in the dragged Brownian particle model treated in this paper.  

   For the pendulum system, Ref. \cite{DJG06} considered 
the work $\mathcal{W}_{p}$ 
done by the external torque $M_{s}$ on the pendulum ($p$).  
   This work is given there by 
\begin{eqnarray}
   \mathcal{W}_{p} = 
   \int_{t_{0}}^{t} ds\;  (M_{s}-M_{t_{0}}) \frac{d \theta_{s}}{ds} .
\label{WorkTorque1}
\end{eqnarray}
%
%   In Ref. \cite{DJG06} an asymptotic fluctuation theorem for 
%this work $\mathcal{W}_{p}$ was derived for  
%a nonequilibrium steady state. 
   Using Eq. (\ref{LinearForci1}) and the correspondences 
in Table \ref{CorreModel1}, this work corresponds to a 
quantity for our dragged Brownian particle model, viz. 
\begin{eqnarray}
   \mathcal{W}_{p} &\longleftrightarrow&
   \int_{t_{0}}^{t} ds \; \kappa v (s-t_{0}) \frac{d x_{s}}{ds} 
   \nonumber\\
   && \spaEq = \mathcal{W}_{l} +\kappa v (t-t_{0}) 
   \left[x_{t} - \frac{1}{2}v(t+t_{0})\right]
   \spaEq 
%\label{}
\end{eqnarray}
which  is clearly different from the works 
$\mathcal{W}_{l}$ and $\mathcal{W}_{c}$ discussed in this paper. 
   In other words, $\mathcal{W}_{l}$, $\mathcal{W}_{c}$ 
and $\mathcal{W}_{p}$ give physically different kinds 
of work in nonequilibrium steady states described by 
a mathematically identical Langevin equation in a dynamical sense.   
   We note that our $\mathcal{W}_{l}$ and $\mathcal{W}_{c}$ 
are consequences of the generalized Onsager-Machlup theory 
in Ref. \cite{TC07a}. 
   We reserve a general discussion on 
fluctuation theorems for different kinds of work for 
a future publication.  

   3) As another nonequilibrium model described by a 
linear Langevin equation, Ref. \cite{ZCC04} considered 
electric circuit models. 
   In that case the system is described by 
a first-order linear Langevin equation,  
which has the same form as the over-damped Langevin equation 
for the dragged Brownian particle model. 
   As a generalization of these electric circuit models, 
an inertial effect in the electric circuit can be  
introduced by including its self-induction. 
   A generalization of the arguments of Ref. \cite{ZCC04} 
to the case including the self-induction, 
as well as a discussion of the effects of self-induction 
on the nonequilibrium work (and heat) fluctuations, 
will be addressed in a future paper.  
   Especially, it would be interesting to observe 
whether there is a critical value of the 
self-induction, above which similar oscillatory effects occur, 
as appear above the critical mass in the inertial case in this paper.

   4) The critical mass $m^{*}$
discussed in this paper for work fluctuations 
also appears in the dynamics 
of the average position $\langle x_{s}\rangle$. 
   In order to discuss this point, 
we note that taking the average of Eq. (\ref{LangeEquat1}),  
the average position $\langle x_{s}\rangle$ of the particle 
at time $s$ satisfies  
\begin{eqnarray}
   m\frac{d^{2} \langle x_{s}\rangle}{ds^{2}} 
   = - \alpha \frac{d \langle x_{s}\rangle}{ds} 
     - \kappa \left(\langle x_{s}\rangle-vs\right)  
\label{LangeAvera1}
\end{eqnarray}
using $\langle \zeta_{s}\rangle =0$. 
   Using $\nus{\pm}$  
defined by Eq. (\ref{CoeffNuPM}),
the solution of Eq. (\ref{LangeAvera1}) is given by 
\begin{eqnarray}
   \langle x_{s}\rangle = v(s-\tau_{r}) 
   + C^{\prime} e^{-\nus{+}s} 
   + C^{\prime\prime} e^{-\nus{-}s}  
\label{LangeAvera2}
\end{eqnarray}
where the constants $C^{\prime}$ and $C^{\prime\prime}$ are 
determined by the average initial conditions  
$\langle x_{t_{0}} \rangle$ and $\langle \dot{x}_{t_{0}}\rangle$   
and are given by 
\begin{eqnarray}
   C^{\prime} &=& 
      -\frac{\nus{-}e^{\nus{+}t_{0}}}{\nus{+}-\nus{-}} 
      \left[\langle x_{t_{0}}\rangle - v(t_{0}-\tau_{r}) \right] 
      \nonumber \\
   &&\spaEq 
      -\frac{e^{\nus{+}t_{0}}}{\nus{+}-\nus{-}} 
      \left(\langle \dot{x}_{t_{0}}\rangle - v \right) , \\
   C^{\prime\prime} &=& 
      \frac{\nus{+}e^{\nus{-}t_{0}}}{\nus{+}-\nus{-}} 
      \left[\langle x_{t_{0}}\rangle - v(t_{0}-\tau_{r}) \right] 
      \nonumber \\
   &&\spaEq 
      +\frac{e^{\nus{-}t_{0}}}{\nus{+}-\nus{-}} 
      \left(\langle \dot{x}_{t_{0}}\rangle - v \right) . 
%\label{}
\end{eqnarray}
   Since the $\nus{\pm}$ include nonzero imaginary parts 
for $m>m^{*}$,  a time-oscillatory behavior 
appears in the average position $\langle x_{s}\rangle$ 
for masses above this critical mass $m^{*}$. 
   This kind of phenomenon   
was discussed for a damped oscillator model \cite{LL69}, 
but its effect on fluctuations 
%of nonequilibrium thermodynamical quantities, like the work, 
in a nonequilibrium steady state   
has not been discussed to the best of our knowledge. 

   In Ref. \cite{TC07a}, we discussed that 
in the over-damped case, the most probable path, 
which is a solution of the Euler-Lagrange equation 
for the Lagrangian function in the Onsager-Machlup theory, 
is expressed as a combination of forward and backward paths.  
   This is also true  
in the inertial case, in which the most probable path 
is given by a solution of the ``Euler-Lagrange'' 
equation (\ref{LagraEquat1}) for $\lambda=0$. 
   To show this, we note that the exponentially decaying terms 
$\exp(-\nus{+}s)$ and $\exp(-\nus{-}s)$ 
on the right-hand side of Eq. (\ref{LangeAvera2}) 
refer to a forward path. 
   We can also introduce the corresponding backward path, 
as a combination of exponentially divergent terms 
$\exp(\nus{+}s)$ and $\exp(\nus{-}s)$. 
   A combination of these forward and backward paths 
gives then the most probable path $\{x_{s}^{*}\}_{s\in [t_{0},t]}$
%the solution (\ref{TildeFunctX1}) of the ``Euler-Lagrange'' 
%equation (\ref{LagraEquat1}) 
for $\lambda=0$, i.e. Eq. (\ref{TildeFunctX1}). 
%   In Ref. \cite{TC07a}, this property was 
%used to discuss the most probable path 
%like $\{x_{s}^{*}\}_{s\in [t_{0},t]}$ for $\lambda=0$ 
%in the Onsager-Machlup theory in the over-damped case.  

   5) There is still the open question of an  
analytical discussion of the asymptotic form 
of $\Delta G(t)$ with the time-oscillations 
shown in Figs. \ref{fig3oscilG0A}, \ref{fig4oscilG0B}, 
\ref{fig6oscilG1A} and  \ref{fig7oscilG1B}.  
%   In this paper we derived an asymptotic form 
%(\ref{FunctGAsym1}) of $G(t)$ analytically, 
%but we did not derive analytically an asymptotic form 
%of $\Delta G(t)$ as the deviation (\ref{DeltaG1}) 
%of $G(t)$ from Eq. (\ref{FunctGAsym1}). 
   In this paper we only analyzed $\Delta G(t)$ numerically 
by fitting it to the function (\ref{FitFunct1}),    
but in principle, such analytical information 
on $\Delta G(t)$ is contained  
in the general form (\ref{FunctG2}) of $G(t)$. 

   6) We have considered the asymptotic 
fluctuation theorem for work in this paper. 
   We now address very briefly its connection with 
other fluctuation theorems. 
   
   (6a) One of the other fluctuation theorems is the 
transient fluctuation theorem \cite{ES94}.  
    This fluctuation theorem was already derived 
and discussed for a dragged Brownian 
particle model with inertia in Ref. \cite{TC07a}.  
   There, we derived transient fluctuation theorems,  
not only for the same works as those in this paper,  
but also for an energy loss by friction. 
   Different from the work, the distribution function for 
the energy loss by friction does \emph{not} satisfy an asymptotic 
fluctuation theorem. 
%   In this sense, different from transient fluctuation theorems, 
%the general conditions with which an asymptotic fluctuation theorem 
%like Eq. (\ref{AsympFT1}) can be obtained is an open question.  

   (6b) Another important fluctuation theorem is 
the extended heat fluctuation theorem \cite{ZC03a,BGG06}.    
   In Ref. \cite{TC07a} we gave a simple derivation 
of this fluctuation theorem, 
based on the assumptions that 
(A) a correlation between the work and the energy difference 
at time $t$ (as well as a correlation between the energies 
at the initial time $t_{0}$ and the final time $t$) 
disappears in the long time limit $t\rightarrow +\infty$, 
(B) the work satisfies the asymptotic fluctuation theorem, 
(C) the work distribution function approaches a Gaussian 
distribution asymptotically in time, and 
(D) the distribution function $P_{e}(E)$ for energy $E$ 
is canonical-like, namely 
$P_{e}(E) \approx \exp(-\beta E)$ for $E>0$. 
   The same derivation could be applied to all models which satisfy 
these four conditions (A), (B), (C) and (D).  
   In particular, using this derivation, 
one can derive an analytical expression for the asymptotic 
heat distribution function itself, as well as  
the extended heat fluctuation theorem not only for 
the over-damped case, as was done in Ref. \cite{TC07a}, 
but also for the inertial case.

%%%%%%%%%%%%%%%%%%%%%%%%%%%%%%%%%%%%%%%%%%%%%%%%%%%%%%%%%%%%%%%%%%%%%%
\section*{Acknowledgements}

   We gratefully acknowledge financial support 
of the National Science Foundation, under award PHY-0501315.

%%%%%%%%%%%%%%%%%%%%%%%%%%%%%%%%%%%%%%%%%%%%%%%%%%%%%%%%%%%%%%%%%%%%%%
%%%%%%%%%%%%%%%%%%%%%%%%%%%%%%%%%%%%%%%%%%%%%%%%%%%%%%%%%%%%%%%%%%%%%%
%\pagebreak
\appendix
\setcounter{section}{0} 
\makeatletter 
   \@addtoreset{equation}{section} 
   \makeatother 
   \def\theequation{\Alph{section}.% 
   \arabic{equation}} 
   
\section{Asymptotic Property for the Matrix $\Lambda_{t}$}
\label{AsympWork}

   In this Appendix we prove Eq. (\ref{AsympCondi1}) 
for the matrix $\Lambda_{t}$.  
   To show this equation in a simple way, without losing 
generality we take the origin of time at $(t_{0}+t)/2$ so 
that the initial time is given by $t_{0}=-t$, 
only in this Appendix.  

   To consider the structure of the matrix $\Lambda_{t}$ 
defined by Eq. (\ref{MatriLambd1}) 
in the long time limit $t\rightarrow+\infty$, we first 
calculate the asymptotic form of  
the matrix $ \Gamma \Phi_{t} \Gamma$, which is an 
essential element of the matrix $\Lambda_{t}$. 
   For this purpose we note 
\begin{widetext}
\begin{eqnarray}
   \mathbf{K}_{s} \mathbf{K}_{s}^{T} 
      = \left(\begin{array}{cccc}
      e^{2\nus{+}s}  
         & e^{(\nus{+}+\nus{-}) s} 
         & e^{(\nus{+}-\nus{-}) s} 
         & 1 \\
      e^{(\nus{+}+\nus{-}) s} 
         & e^{2\nus{-} s}  
         & 1 
         & e^{-(\nus{+}-\nus{-}) s} \\
      e^{(\nus{+}-\nus{-})  s}  
         & 1 
         & e^{-2\nus{-} s} 
         & e^{-(\nus{+}+\nus{-}) s} \\
      1  
         & e^{-(\nus{+}-\nus{-}) s} 
         & e^{-(\nus{+}+\nus{-}) s} 
         & e^{-2\nus{+} s} 
   \end{array}\right) .
\label{MatriKKT1}
\end{eqnarray}
   Inserting Eq. (\ref{MatriKKT1}) into Eq. (\ref{FunctPhi1}) 
and using the relation $t_{0}=-t$ we obtain 
\begin{eqnarray}
   \Phi_{t} =   2 \left(\begin{array}{cccc}
      \frac{\sinh(2\nus{+}t)}{2\nus{+}}  
         & \frac{\sinh[(\nus{+}+\nus{-}) t]}{\nus{+}+\nus{-}} 
         & \frac{\sinh[(\nus{+}-\nus{-}) t]}{\nus{+}-\nus{-}} 
         & t \\
      \frac{\sinh[(\nus{+}+\nus{-}) t]}{\nus{+}+\nus{-}} 
         & \frac{\sinh(2\nus{-} t)}{2\nus{-}}  
         & t 
         & \frac{\sinh[(\nus{+}-\nus{-}) t]}{\nus{+}-\nus{-}} \\
      \frac{\sinh[(\nus{+}-\nus{-})  t]}{\nus{+}-\nus{-}}  
         & t 
         & \frac{\sinh(2\nus{-} t)}{2\nus{-}} 
         & \frac{\sinh[(\nus{+}+\nus{-})t]}{\nus{+}+\nus{-}} \\
      t  
         & \frac{\sinh[(\nus{+}-\nus{-}) t]}{\nus{+}-\nus{-}} 
         & \frac{\sinh[(\nus{+}+\nus{-}) t]}{\nus{+}+\nus{-}} 
         & \frac{\sinh(2\nus{+} t)}{2\nus{+}} 
   \end{array}\right) 
\label{MatriPhi2}
\end{eqnarray}
with the hyperbolic function $\sinh(x)\equiv 
[\exp(x)-\exp(-x)]/2$.  
   Equations (\ref{functGamma1}) and (\ref{MatriPhi2}) lead to 
\begin{eqnarray}
   \Gamma \Phi_{t} \Gamma
   = \left(\begin{array}{cc}
         \Psi_{t} & 0_{2} \\ 0_{2} & 0_{2}
      \end{array}\right) 
\label{MatriSuppl1}
\end{eqnarray}
where $0_{2}$ is the $2\times 2$ null matrix, and 
the $2\times 2$ matrix $\Psi_{t}$ is given by 
\begin{eqnarray}
   \Psi_{t}&\equiv&  
      4\left(\begin{array}{cc}
         \nus{+} \sinh(2\nus{+}t) 
         & \frac{2\nus{+}\nus{-}}{\nus{+}+\nus{-}} 
            \sinh[(\nus{+}+\nus{-}) t]  \\
         \frac{2\nus{+}\nus{-}}{\nus{+}+\nus{-}}
             \sinh[(\nus{+}+\nus{-}) t] 
         & \nus{-}\sinh(2\nus{-} t)     
      \end{array}\right) 
      \nonumber \\
   &\stackrel{t\rightarrow+\infty}{\sim}& 
      2 \left(\begin{array}{cc}
          \nus{+} e^{2\nus{+}t} 
         & \frac{2\nus{+}\nus{-}}{\nus{+}+\nus{-}}
               e^{(\nus{+}+\nus{-}) t}\\
         \frac{2\nus{+}\nus{-}}{\nus{+}+\nus{-}} 
            e^{(\nus{+}+\nus{-}) t} 
         & \nus{-}  e^{2\nus{-} t}     
      \end{array}\right) 
      \nonumber \\
   &=& 
      2 \left(\begin{array}{cc}
         e^{\nus{+}t} & 0 \\ 0 & e^{\nus{-}t} 
      \end{array}\right) 
      \left(\begin{array}{cc}
         \nus{+} 
         & \frac{2\nus{+}\nus{-}}{\nus{+}+\nus{-}}   \\
         \frac{2\nus{+}\nus{-}}{\nus{+}+\nus{-}}
         & \nus{-}    
      \end{array}\right) 
      \left(\begin{array}{cc}
         e^{\nus{+}t} & 0 \\ 0 & e^{\nus{-}t} 
      \end{array}\right) . \;\;\;\;\;\;\;\;\;
      \label{MatriPsi1} 
\end{eqnarray}
\end{widetext}
   Here, we used the positivity $Re\{\nus{\pm}\}>0$ 
of the real part of $\nus{\pm}$  
(assuming a nonzero mass $m \neq 0$ and 
a nonzero spring constant $\kappa \neq 0$) and also 
$\sinh(at) \stackrel{t\rightarrow+\infty}{\sim} 
(1/2)\exp(at)$ for any number $a$ with the positive real part  
$Re\{a\}>0$. 
%   [Here, $Re\{a\}$ is the real part of any complex number $a$.]

   Second, we obtain a simplified form of the 
matrix $A_{t}^{-1}$  in the long time limit, 
which is another essential element 
of the matrix $\Lambda_{t}$. 
   Noting again that the real part of the number 
$\nus{\pm}$ is (strictly non-zero) positive and the initial 
time is given by $t_{0}=-t$, 
we obtain the asymptotic form of the matrix 
$A_{t}$ defined by Eq. (\ref{MatriA1}) as 
\begin{eqnarray}
   A_{t} &\stackrel{t\rightarrow+\infty}{\sim}& 
   \left(\begin{array}{cc}0_{2} & A_{t}^{(1)} \\
    A_{t}^{(2)} & 0_{2}\end{array}\right) . 
\label{MatriA3}
\end{eqnarray} 
for the long time limit $t\rightarrow+\infty$.  
   Here, $0_{2}$ is the $2\times2$ null matrix, and 
$A_{t}^{(j)}$, $j=1,2$ are defined by 
\begin{eqnarray}
   A_{t}^{(1)} &\equiv&  
   \left(\begin{array}{cc}
      e^{\nus{-}t} & e^{\nus{+}t}  \\
      -\nus{-}e^{\nus{-}t} & 
         -\nus{+}e^{\nus{+}t}  
    \end{array}\right) ,
    \label{SubMatrixA1} \\
   A_{t}^{(2)} &\equiv& 
   \left(\begin{array}{cc}
      e^{\nus{+}t} & e^{\nus{-}t}  \\
       \nus{+}e^{\nus{+}t} & 
         \nus{-}e^{\nus{-}t}
    \end{array}\right) .
%\label{}
\end{eqnarray}
   From Eq. (\ref{MatriA3}) we derive 
\begin{eqnarray}
   A_{t}^{-1} \stackrel{t\rightarrow+\infty}{\sim} 
   \left(\begin{array}{cc}0_{2} & A_{t}^{(2)}{}^{-1} \\
    A_{t}^{(1)}{}^{-1} & 0_{2}\end{array}\right) 
\label{MatriAInv0}
\end{eqnarray}
where $A_{t}^{(1)}{}^{-1}$ and $A_{t}^{(2)}{}^{-1}$ are given by 
\begin{eqnarray}
   A_{t}^{(1)}{}^{-1} 
   &=&  \frac{1}{\nus{-}-\nus{+}} 
      \left(\begin{array}{cc}
         -\nus{+}e^{-\nus{-}t}  & - e^{-\nus{-}t}     \\
      \nus{-}e^{-\nus{+}t} & e^{-\nus{+}t}
      \end{array}\right) ,
      \label{MatriAInv1}
   \\
   A_{t}^{(2)}{}^{-1} 
   &=&  \frac{1}{\nus{-}-\nus{+}}
      \left(\begin{array}{cc}
        \nus{-}e^{-\nus{+}t}  & - e^{-\nus{+}t}     \\
        - \nus{+}e^{-\nus{-}t} & e^{-\nus{-}t}
      \end{array}\right) .
   \label{MatriAInv2}
\end{eqnarray}
Eq. (\ref{MatriAInv0}) give an asymptotic form for the matrix 
$A_{t}^{-1}$. 

   Finally, using Eqs. (\ref{MatriLambd1}), (\ref{MatriSuppl1})  
and (\ref{MatriAInv0}) we obtain the asymptotic form 
of the matrix $\Lambda_{t}$ as  
\begin{eqnarray}
   \Lambda_{t} 
      &\stackrel{t\rightarrow+\infty}{\sim}& 
       \left(\begin{array}{cc}
            0_{2}  & 0_{2}  \\  0_{2} 
           &  \left[A_{t}^{(2)}{}^{-1}\right]^{T}
              \Psi_{t} A_{t}^{(2)}{}^{-1} 
        \end{array}\right) .
\label{MatriLambd3}
\end{eqnarray}
   Here, using Eqs. (\ref{MatriPsi1})  
and (\ref{MatriAInv2}) 
the non-vanishing matrix elements of the matrix 
(\ref{MatriLambd3}) is given by  
\begin{eqnarray}
   \left[A_{t}^{(2)}{}^{-1}\right]^{T}\Psi_{t} A_{t}^{(2)}{}^{-1} 
   &\;\stackrel{t\rightarrow+\infty}{\sim}\;&
      2 \left(\begin{array}{cc}  
         \frac{\nus{+}\nus{-}}{\nus{+}+\nus{-}}& 0\\
         0 &\frac{1}{\nus{+}+\nus{-}}
      \end{array}\right)
      \nonumber \\
   &=& \frac{2}{\alpha} \left(\begin{array}{cc}  
         \kappa & 0 \\
         0 & m
      \end{array}\right)
\label{AsympLambd1}
\end{eqnarray}
where we used $\nus{+}+\nus{-} = \alpha/m$ and 
$\nus{+}\nus{-} = \kappa/m$. 
   By Eqs. (\ref{MatriLambd3}) and (\ref{AsympLambd1}) we obtain
\begin{eqnarray}
   \lim_{t\rightarrow +\infty}\Lambda_{t} =
      \frac{2}{\alpha} \left(\begin{array}{cccc}
         0&0&0&0\\
         0&0&0&0\\
         0&0& \kappa & 0\\
         0&0&0& m
      \end{array}\right) .
\label{MatriLambd4}
\end{eqnarray}
   Equation (\ref{MatriLambd4}) shows that the 
matrix $\Lambda_{t}$ approaches a time-independent constant 
matrix in the long time limit $t\rightarrow +\infty$. 
   Therefore, the matrix $\Lambda_{t}/(t-t_{0})$ approaches  
the $4\times 4$ null matrix in the long time 
limit $t\rightarrow +\infty$, implying that the condition 
(\ref{AsympCondi1}) is satisfied.

%%%%%%%%%%%%%%%%%%%%%%%%%%%%%%%%%%%%%%%%%%%%%%%%%%%%%%%%%%%%%%%%%%%%%%
\section{Work Distribution for the Nonequilibrium Steady State}
\label{FinitWork}

   In this Appendix we give a derivation of Eq. (\ref{DistriWork9}) 
for the work distribution function $P(W,t)$ in the case of the 
nonequilibrium steady state initial condition (\ref{InitiNNSS1}). 

   First, we note that the initial distribution 
function (\ref{InitiNNSS1}) can be written in the form  
\begin{eqnarray}
   f(x_{i},p_{i},t_{0}) 
   = \frac{\beta}{2\pi} \sqrt{\frac{\kappa}{m}} 
   \exp\left\{-\frac{\alpha\beta}{4}
   \left[\mathbf{B}_{if}^{(1)}\right]^{T} \Lambda^{(0)}
   \mathbf{B}_{if}^{(1)}\right\} , 
   \nonumber\\ 
\label{InitiNNSS2}
\end{eqnarray}
using Eqs. (\ref{VectoBJ}) and (\ref{TildeLambda1}).  
   Equation (\ref{InitiNNSS2}) means that 
the initial distribution function $f(x_{i},p_{i},t_{0})$ is 
Gaussian for the components of the vector 
$\mathbf{B}_{if}^{(1)}$. 
   Using Eq. (\ref{InitiNNSS2}) and again the vector 
$\mathbf{B}_{if}^{(1)}$ given by Eq. (\ref{VectoBJ}), 
the work distribution function (\ref{DistriWork6}) 
can be represented by 
\begin{widetext}
\begin{eqnarray}
   P_{w}(W,t) &=&  
      \frac{C_{\mathcal{E}} m \beta}{4\pi v} 
      \sqrt{\frac{\kappa m}{
      \pi\alpha\beta \left( t-t_{0} 
      -\tau_{r}^{2}\mathbf{J}^{T}\Lambda_{t}\mathbf{J}\right)}} 
       %\int\!\int dx_{i}dp_{i} \int\!\int dx_{f}dp_{f} 
       \int\!\int\!\int\!\int d \mathbf{B}_{if}^{(1)}
      \nonumber \\
   && \spaEq\times
       \exp\left[
       -\frac{1}{4}\alpha\beta \left[\mathbf{B}_{if}^{(1)}\right]^{T}
      \left( \Lambda^{(0)} 
      + \Lambda_{t} \right)\mathbf{B}_{if}^{(1)}
  %    \right]
  %    \nonumber \\
  % && \times
  %    \exp\left\{ 
      - \frac{\left[ 
      \alpha\beta v 
      \left(\bfeta^{T}-\tau_{r}\mathbf{J}^{T}\Lambda_{t}\right)
      \mathbf{B}_{if}^{(1)} 
      - W +\alpha\beta v^{2}(t-t_{0})\right]^{2}}
      {4\alpha\beta v^{2} \left( t-t_{0} 
      -\tau_{r}^{2}\mathbf{J}^{T}\Lambda_{t}\mathbf{J}\right)}\right\}
      \spaEq 
     % \nonumber \\&&
\label{DistriWork6b}
\end{eqnarray}
where we used the relation $dx_{i}dp_{i} dx_{f}dp_{f} =
m^{2} d \mathbf{B}_{if}^{(1)}$ 
because of Eq. (\ref{VectoBJ}). 

   Now, we note that 
\begin{eqnarray}
    (\tilde{\bfa}^{T} \tilde{\bfx} + \tilde{b})^{2} 
   + \tilde{\bfx}^{T} \tilde{\mathcal{C}} \tilde{\bfx} 
   %   \nonumber \\
   &=& \tilde{\bfx}^{T}\left(\tilde{\bfa} 
      \tilde{\bfa}^{T}+\tilde{\mathcal{C}}\right)\tilde{\bfx} 
      + 2\tilde{b}\tilde{\bfa}^{T}\tilde{\bfx} +\tilde{b}^{2}   
      \nonumber \\
   &=& \left[\tilde{\bfx}+\tilde{b}
      \left(\tilde{\bfa} \tilde{\bfa}^{T}
      +\tilde{\mathcal{C}}\right)^{-1}
      \tilde{\bfa}\right]^{T}
      \left(\tilde{\bfa} \tilde{\bfa}^{T}+\tilde{\mathcal{C}}\right)
      \left[\tilde{\bfx}+\tilde{b}\left(\tilde{\bfa} \tilde{\bfa}^{T}
      +\tilde{\mathcal{C}}\right)^{-1}
      \tilde{\bfa}\right] 
  %    \nonumber \\
  % &&\spaEq\spaEq 
      +\tilde{b}^{2} \left[1- \tilde{\bfa}^{T}
      \left(\tilde{\bfa} \tilde{\bfa}^{T}
      +\tilde{\mathcal{C}}\right)^{-1}\tilde{\bfa} \right] 
      \nonumber \\
\label{SquarSuppl1}
\end{eqnarray}
for any $n$-dimensional vector $\tilde{\bfa}$ and $\tilde{\bfx}$, any scalar $\tilde{b}$,  
and any $n\times n$ symmetric matrix $\tilde{\mathcal{C}}$ existing 
the inverse matrix of $\tilde{\bfa} \tilde{\bfa}^{T}
+\tilde{\mathcal{C}}$. 
   Applying Eq. (\ref{SquarSuppl1}) to Eq. (\ref{DistriWork6b}) 
for the case of $\tilde{\bfx} = \mathbf{B}_{if}^{(1)}$, 
$\tilde{\bfa} = \alpha\beta v
(\bfeta-\tau_{r}\Lambda_{t}\mathbf{J})$, 
$\tilde{b} = \alpha\beta v^{2}(t-t_{0}) - W$ and 
$ \tilde{\mathcal{C}} = (\alpha\beta v)^{2} ( t-t_{0} 
      -\tau_{r}^{2}\mathbf{J}^{T}\Lambda_{t}\mathbf{J}) 
      (  \Lambda^{(0)} +\Lambda_{t})$, 
and carrying out the integral over $\mathbf{B}_{if}^{(1)} (=\tilde{\bfx})$ 
in Eq. (\ref{DistriWork6b}), we obtain 
\begin{eqnarray}
   P_{w}(W,t)   
   &=& 
      \frac{C_{\mathcal{E}} m \beta}{4\pi v} 
      \sqrt{\frac{\kappa m}{
      \pi\alpha\beta \left( t-t_{0} 
      -\tau_{r}^{2}\mathbf{J}^{T}\Lambda_{t}\mathbf{J}\right)}} 
   %  \nonumber \\
   %&&\spaEq  \times
       \int\!\int\!\int\!\int d\tilde{\bfx} \; 
      \exp\left\{ 
      - \frac{(\tilde{\bfa}^{T} \tilde{\bfx} + \tilde{b})^{2} 
      + \tilde{\bfx}^{T} \tilde{\mathcal{C}} \tilde{\bfx} }
      {4\alpha\beta v^{2} \left( t-t_{0} 
      -\tau_{r}^{2}\mathbf{J}^{T}\Lambda_{t}\mathbf{J}\right)}\right\}
      \nonumber \\
   &=& C_{w}^{\prime}\exp\left\{ -\frac{
      \left[1- 
      \tilde{\bfa}^{T}\left(\tilde{\bfa} \tilde{\bfa}^{T}
      +\tilde{\mathcal{C}}
      \right)^{-1}\tilde{\bfa} \right] }
      {4\alpha\beta v^{2} \left( t-t_{0} 
      -\tau_{r}^{2}\mathbf{J}^{T}\Lambda_{t}\mathbf{J}\right)}
       \tilde{b}^{2} \right\}
      \nonumber \\
   &=& C_{w}^{\prime}\exp\left\{ -\frac{
      1- \Omega_{t}  }
      {4\alpha\beta v^{2} \left( t-t_{0} 
      -\tau_{r}^{2}\mathbf{J}^{T}\Lambda_{t}\mathbf{J}\right)}
      \left[ W -\alpha\beta v^{2}(t-t_{0}) \right]^{2}\right\}
      %\nonumber\\ 
\label{DistriWork8}
\end{eqnarray}
\end{widetext}
using a normalization constant $C_{w}^{\prime}$ and 
Eq. (\ref{FunctOmega1}) for $\Omega_{t}$.  
   The constant $C_{w}^{\prime}$ in Eq. (\ref{DistriWork8}) 
can be determined by 
the normalization condition $\int dW\; P(W,t) = 1$, 
which and Eq. (\ref{DistriWork8}) yield Eq. (\ref{DistriWork9}).

%%%%%%%%%%%%%%%%%%%%%%%%%%%%%%%%%%%%%%%%%%%%%%%%%%%%%%%%%%%%%%%%%%%%%%
\section{Asymptotic Form of $G(t)$}
\label{AsympFunctG} 

   In this Appendix we give an argument to derive 
Eq. (\ref{FunctGAsym1}) for $G(t)$. 

   The essential point to derive 
Eq. (\ref{FunctGAsym1}) for $G(t)$ is the 
asymptotic form (\ref{MatriLambd4}), or equivalently 
\begin{eqnarray}
   \lim_{t\rightarrow +\infty}\Lambda_{t} =
      2 \left(\begin{array}{cccc}
         0&0&0&0\\
         0&0&0&0\\
         0&0&1/\tau_{r} & 0 \\
         0&0&0& \tau_{m}
      \end{array}\right) 
\label{MatriLambd5}
\end{eqnarray}
using the relations $\tau_{r}=\alpha/\kappa$ and $\tau_{m}=m/\alpha$. 
   Using Eqs. (\ref{MatriVarpi1}), (\ref{VectorJ1}),  
(\ref{TildeLambda1}) and (\ref{MatriLambd5}) we obtain 
\begin{eqnarray}
   \lim_{t\rightarrow +\infty}
      \mathbf{J}^{T}\Lambda_{t}\mathbf{J} &=& 2/\tau_{r} ,
      \label{LambdJLJ1} 
      \\
   \lim_{t\rightarrow +\infty}
      \left(\bfeta-\tau_{r}\Lambda_{t}\mathbf{J} \right)
      &=& 
      \left(\begin{array}{c}
         -1\\ - \tau_{m}\vartheta \\ 
         -1\\ \tau_{m}\vartheta
      \end{array}\right) ,
      \label{MatriVarpi2} \\
   \lim_{t\rightarrow +\infty}
      (\Lambda^{(0)}+\Lambda_{t}) &=& 
      2 \left(\begin{array}{cccc}
         1/\tau_{r} &0&0&0\\
         0& \tau_{m} &0&0\\
         0&0&1/\tau_{r} & 0 \\
         0&0&0& \tau_{m}
      \end{array}\right) .
\label{MatriLambd6}
\end{eqnarray}
   Equations (\ref{LambdJLJ1}), (\ref{MatriVarpi2}) and 
(\ref{MatriLambd6}) lead to 
\begin{widetext}
\begin{eqnarray}
   &&\left[\left(\bfeta-\tau_{r}\Lambda_{t}\mathbf{J}\right)
          \left(\bfeta-\tau_{r}\Lambda_{t}\mathbf{J}\right)^{T}
      +\left( t-t_{0} 
         -\tau_{r}^{2}\mathbf{J}^{T}\Lambda_{t}\mathbf{J}\right) 
         \left( \Lambda^{(0)}+\Lambda_{t} \right) \right]^{-1} 
         \nonumber \\
%   &&\spaEq  \stackrel{t\rightarrow+\infty}{\sim} 
%      \frac{\tau_{r}}{4 (t-t_{0}- 2\tau_{r}) 
%      \left[t-t_{0}-\tau_{r}+\tau_{m}\vartheta^2\right]}
%         \nonumber \\
%   &&\spaEq\spaEq\spaEq \times 
%      \left(\begin{array}{cc}
%         2(t-t_{0})-3\tau_{r}+2\tau_{m}\vartheta^{2} 
%         & -\vartheta \\
%         -\vartheta & 
%            \frac{2(t-t_{0}-\tau_{r})+\tau_{m}\vartheta^{2}}
%               {\tau_{r}\tau_{m}} \\
%         - \tau_{r} &   -\vartheta \\
%         \vartheta & \frac{\vartheta^2}{\tau_{r}} 
%   \end{array}\right.
%         \nonumber \\
%   &&\spaEq\spaEq\spaEq\spaEq\spaEq 
%      \left.\begin{array}{cccc}
%         -\tau_{r}  & \vartheta \\
%         -\vartheta & \frac{\vartheta^2}{\tau_{r}}\\
%         2(t-t_{0})-3\tau_{r}+2\tau_{m}\vartheta^{2} 
%            & \vartheta \\
%         \vartheta  
%           &\frac{2(t-t_{0}-\tau_{r})+\tau_{m}\vartheta^{2}}
%              {\tau_{r}\tau_{m}}
%   \end{array}\right) . \;\;\;\;\;\;
%         \nonumber \\
   &&  \spaEq\stackrel{t\rightarrow+\infty}{\sim} 
      \frac{\tau_{r}}{2 (t-t_{0}- 2\tau_{r}) \Xi_{t}}
   %      \nonumber \\
   %&&\spaEq\spaEq\spaEq \times 
      \left(\begin{array}{cccc}
         \Xi_{t}-\tau_{r} & -\vartheta 
            & -\tau_{r} & \vartheta \\
         -\vartheta 
            &\frac{\Xi_{t}-\tau_{m}\vartheta^{2}}{\tau_{r}\tau_{m}} 
            &-\vartheta & \frac{\vartheta^2}{\tau_{r}}\\
         - \tau_{r} &   -\vartheta 
            &\Xi_{t}-\tau_{r} & \vartheta \\
         \vartheta & \frac{\vartheta^2}{\tau_{r}} 
            &\vartheta 
            & \frac{\Xi_{t}-\tau_{m}\vartheta^{2}}{\tau_{r}\tau_{m}}
   \end{array}\right) \;\;\;\;\;\;
   \nonumber \\
\label{InverLambd1}
\end{eqnarray}
\end{widetext}
with $\Xi_{t} \equiv 2\left(t-t_{0}-\tau_{r}+\vartheta^{2}\tau_{m}\right)$. 
   By Eqs. (\ref{FunctOmega1}), (\ref{MatriVarpi2}) 
and (\ref{InverLambd1}), we obtain %
\begin{eqnarray}
   \Omega_{t} 
   \stackrel{t\rightarrow+\infty}{\sim}
   \frac{\tau_{r} + \tau_{m}\vartheta^{2}}{t-t_{0} 
   - \tau_{r} + \tau_{m}\vartheta^{2}} .
\label{AsympOmega1}
\end{eqnarray}
   From Eqs. (\ref{FunctG2}), (\ref{LambdJLJ1}) 
and (\ref{AsympOmega1}) we derive Eq. (\ref{FunctGAsym1}).

%%%%%%%%%%%%%%%%%%%%%%%%%%%%%%%%%%%%%%%%%%%%%%%%%%%%%%%%%%%%%%%%%%%%%%

%%%%%%%%%%%%%%%%%%%%%%%%%%%%%%%%%%%%%%%%%%%%%%%%%%%%%%%%%%%%%%%%%%%%%%

\end{document}